\documentclass[structabstract]{aa}  
\usepackage{graphicx}
\usepackage{natbib}
\usepackage{lscape}
\usepackage{longtable}
\usepackage{txfonts}
\def\macc   {$\dot{M}_{\rm acc}$}
\def\maccnoise   {$\dot{M}_{\rm acc,noise}$}
\def\lacc   {$L_{\rm acc}$}
\def\laccnoise   {$L_{\rm acc,noise}$}
\def\lline {$L_{\rm line}$}

\begin{document}
   \title{X-Shooter spectroscopy of young stellar objects:\thanks{Based on observations collected in the programs 084.C-0269, 085.C-0238, 086.C-0173, 087.C-0244, 089.C-0143 at the European Organisation for Astronomical Research in the Southern Hemisphere (Chile).}\\ II. Impact of chromospheric emission on accretion rate estimates}

   \author{C.F. Manara\inst{1}, L. Testi\inst{1,2}, E. Rigliaco\inst{3},  J.M. Alcala\inst{4}, A. Natta\inst{2,5}, B. Stelzer\inst{6}, \\ K. Biazzo\inst{4}, E. Covino\inst{4}, S. Covino\inst{7}, G. Cupani\inst{8}, V. D'Elia\inst{9}, S. Randich\inst{2}           }

   \institute{European Southern Observatory, Karl Schwarzschild Str. 2, 85748 Garching, Germany\\
              \email{cmanara@eso.org}
         \and
             INAF - Osservatorio Astrofisico di Arcetri, Largo E.Fermi 5, I-50125 Firenze, Italy
	\and
		Department of Planetary Science, Lunar and Planetary Lab, University of Arizona, 1629, E. University Blvd, 85719, Tucson, AZ
	\and
		INAF - Osservatorio Astronomico di Capodimonte, Via Moiariello 16, I-80131 Napoli, Italy
	\and
	School of Cosmic Physics, Dublin Institute for Advanced Studies, 31 Fitzwilliam Place, Dublin 2, Ireland
	\and
		INAF - Osservatorio Astronomico di Palermo, Piazza del Parlamento 1, I-90134 Palermo, Italy
	\and
		INAF - Osservatorio Astronomico di Brera, Via Bianchi 46, I-23807 Merate (LC), Italy
	\and
		INAF - Osservatorio Astronomico di Trieste, via Tiepolo 11, I-34143 Trieste, Italy
	\and
		INAF - Osservatorio Astronomico di Roma, Via di Frascati 33, I-00040 Monte Porzio Catone (RM), Italy
	}

   \date{Received December 14, 2012; accepted January 10, 2013}

 
  \abstract
   {The lack of knowledge of photospheric parameters and the level of chromospheric activity in young low-mass pre-main sequence stars introduces uncertainties when measuring mass accretion rates in accreting (Class II) Young Stellar Objects. A detailed investigation of the effect of chromospheric emission on the estimates of mass accretion rate in young low-mass stars is still missing. This can be undertaken using  samples of young diskless (Class III) K and M-type stars.  
}
   {Our goal is to measure the chromospheric activity of Class III pre main sequence stars to determine its effect on the estimates of accretion luminosity (\lacc) and mass accretion rate (\macc) in young stellar objects with disks.}
   {Using VLT/X-Shooter spectra we have analyzed a sample of 24 non-accreting young stellar objects of spectral type between K5 and M9.5. We identify the main emission lines normally used as tracers of accretion in Class II objects, and we determine their fluxes in order to estimate the contribution of the chromospheric activity to the line luminosity.}
   {We have used the relationships between line luminosity and accretion luminosity derived in the literature for Class II objects to evaluate the impact of chromospheric activity on the accretion rate measurements. We find that the typical chromospheric activity would bias the derived accretion luminosity by \laccnoise $< 10^{-3} L_\odot$, with a strong dependence with the T$_{\rm eff}$ of the objects. The noise on \macc \ depends on stellar mass and age, and the typical values of log($\dot{M}_{\rm acc,noise}$) range between $\sim-9.2$ to $-11.6 M_\odot$/yr.}
   {Values of \lacc $\lesssim 10^{-3} L_\odot$ obtained in accreting low-mass pre main sequence stars through line luminosity should be treated with caution as the line emission may be dominated by the contribution of chromospheric activity. }


\authorrunning{Manara et al.}
\titlerunning{Impact of chromospheric emission on accretion rate estimate of young stellar objects}
   \maketitle
%

\section{Introduction}

Circumstellar disks are formed as a natural consequence of angular momentum conservation during gravitational collapse of cloud cores \citep[e.g.][]{Shu87}. In the early phases of star formation, the disk allows for the dissipation of angular momentum channeling the accretion of material from the infalling envelope onto the central young stellar object (YSO). At later stages, when the envelope is dissipated, planetary systems form in the disk, while the star-disk interaction continues through the inner disk and the stellar magnetosphere. This phenomenon constrains the final stellar mass build-up \citep[e.g.][]{Hartmann98} and its typical timescales are connected with the timescales on which disks dissipate and planet formation occurs \citep[e.g.][]{Hernandez07,Fedele10,Williams11}. 

Accretion can be observed using typical signatures in the spectra of YSOs, such as the continuum excess in the blue part of the visible spectrum \citep[e.g.][]{Gullbring98} and the prominent optical and infrared emission lines \citep[e.g.][]{Muzerolle98, Natta04, Herczeg08, Rigliaco12}. The measurements used to determine the bolometric accretion luminosity (\lacc) are either direct or indirect. Direct measurements are obtained by measuring the emission in excess of the photospheric one in the Balmer and Paschen continua and adopting a model to correct for the emission at the wavelengths not covered by the observations, and which originates mostly below the U-band threshold \citep[e.g.][]{Valenti93,Gullbring98,Calvet98,Herczeg08,Rigliaco12} or by line profile modelling \citep[e.g.][]{Muzerolle98}. Indirect measurements are obtained using empirical correlations between emission line luminosity ($L_{\rm line}$) and \lacc \ \citep[e.g.][]{Muzerolle98b,Natta04,Natta06,Mohanty05,Rigliaco11a}. 

Measurements of mass accretion rate (\macc) are subjected to many uncertainties, because this quantity depends on \lacc \ and on the mass-to-radius ($M_*/R_*$) ratio. This ratio is normally determined from the position of the object on the HR diagram and a set of evolutionary models. Uncertainties on spectral type (SpT) of the objects affect the determination of effective temperature (T$_{\rm eff}$) while those on the extinction ($A_V$) and the distance mainly affect the estimate of stellar luminosity ($L_*$). It is not trivial to determine those parameters in accreting stars because of accretion shocks on the stellar surface producing veiling in the photospheric lines \citep[e.g.][]{Calvet98} and modifying the photometric colors \citep[e.g.][]{DaRio10}. Moreover, the derivation of \lline, from which \lacc \ is determined, is affected by another stellar property, namely the chromospheric activity of the YSOs \citep{Houdebine96,Franchini98}. Chromospheric line emission is usually small compared to the accretion-powered emission, but it can become important when accretion decreases at later evolutionary stages \citep{Ingleby11} and in lower mass stars, where accretion rates are smaller \citep{Rigliaco12}. Therefore, this is an important source of uncertainty that has not been investigated in detail so far.

Part of the INAF consortium's Guaranteed Time Observations (GTO) of X-Shooter, a broad-band, medium-resolution, high-sensitivity spectrograph mounted on the ESO/VLT, has been allocated for star formation studies, in particular to investigate accretion, outflows, and chromospheric emission in low-mass Class II young stellar and sub-stellar objects \citep{GTO}. The targets observed during the GTO were chosen in nearby (d $<$ 500 pc) star forming regions with low extinction, and with many very low-mass (VLM) YSO (M$_* < 0.2$ M$_\odot$). Generally, those YSOs for which measurements in many photometric bands were available, both in the IR \citep[e.g.][]{Hernandez07,Merin08} and in the visible part of the spectrum \citep[e.g.][]{Merin08,Rigliaco11a} were selected. In order to derive \lacc \ of a given Class II YSO a Class III template of the same SpT as the Class II is needed. Therefore, during this GTO survey 24 Class III targets in the range K5-M9.5 have been observed, providing the first broad-band grid of template spectra for low-mass stars and brown dwarfs (BDs). Since these spectra have a very large wavelength range ($\sim$350 - 2500 nm) covering part of the UV spectrum (UVB), the whole Visible (VIS), and the Near Infrared (NIR), this sample allows us to: determine the stellar parameters of the targets, derive the chromospheric emission line fluxes and luminosities and, hence, determine the implications of chromospheric emission on the indirect accretion estimates in Class II objects. 

The paper is structured as follows. In Sect.~\ref{observ_sec} we discuss the sample selection, the observation strategy and the data reduction procedure. In Sect.~\ref{sec::source_class} we describe how SpTs of our targets have been determined, while in Sect.~\ref{sec::star_param} we derive their main stellar parameters. In Sect.~\ref{sec::lines} we identify the main lines present in the spectra and derive their intensities. In Sect.~\ref{sec::macc} we discuss the implications of the line luminosity found for studies of \macc \ in Class II YSOs. Finally, in Sect.~\ref{sec::conclusion} we summarize our conclusions.


\section{Sample, observations and data reduction}
\label{observ_sec}

\subsection{Sample}
Among the objects observed in the GTO survey, we selected only those that have been classified as Class III objects using {\it Spitzer} photometric data. The sample comprises Class III YSOs in the $\sigma$ Orionis, Lupus III and TW Hya associations. In the end, the number of targets is 24; 13 objects are members of the TW Hya association, 6 of the Lupus III cloud and 5 of the $\sigma$ Orionis region. Their SpTs range between K5 and M9.5 (see Sect.~\ref{sec::source_class}). Three BDs, namely Par-Lup3-1, TWA26, and TWA29, are included in our sample, with SpT M6.5, M9, and M9.5, respectively. Data available from the literature for these objects are reported in Table~\ref{known_parameters_table}.

Six YSOs in our sample are components of three known wide visual binary systems. In all cases we were able to resolve them, given that their separations are always larger than 6\arcsec. 

\subsection{Observations}
All the observations have been made in the slit nodding mode, in order to achieve a good sky subtraction. Different exposure times and slit dimensions were used for different targets, in order to have enough S/N and to avoid saturation. The readout mode used in all the observations was ``100,1x1, hg", while the resolution of our spectra is R = 9100, 5100, and 3300 in the UVB arm for slit 0.5\arcsec, 1.0\arcsec, and 1.6\arcsec, respectively;  R = 17400, 8800, and 5400 in the VIS arm for slit 0.4\arcsec, 0.9\arcsec, and 1.5\arcsec, respectively;  R = 11300, 5600, and 3500 in the NIR arm for slit 0.4\arcsec, 0.9\arcsec, and 1.5\arcsec, respectively. We report in Table~\ref{observations_table} the details of all observations performed for this work.

\subsection{Data reduction}
\label{sec::data_red}
The data reduction has been done using two versions of the X-Shooter pipeline \citep{Modigliani}, run through the {\it EsoRex} tool, according to the period in which the data were acquired: version 1.0.0 has been used for the data of December 2009 and May 2010, while for data gathered in January 2011, April 2011, and April 2012, version 1.3.7 has been used. The two versions lead to results that are very similar. The reduction has been done independently for each spectrograph arm. This takes into account also the flexure compensation and the instrumental profile. We have used the pipeline recipe {\fontfamily{phv}\selectfont xsh\_scired\_slit\_nod} which includes bias and flat-field correction, wavelength calibration, order tracing and merging and flux calibration. Regarding the last point, by comparison of the response function of different flux standards observed during the same night, we estimate an intrinsic error on the flux calibration of less than 5\%. Given that some observations have been done with poor weather conditions (seeing $\sim$3.5\arcsec) or with narrow-slits, we have then checked the flux calibration of each object using the available photometric data, usually in the $U, B, V, R, I, J, H, K$ bands, as reported in Table~\ref{known_parameters_table}. We have verified that all the spectra match well the photometric spectral energy distribution (SED) and adjusted the flux-calibrated spectra to match the photometric flux. Binaries have been reduced in stare mode. Telluric removal has been done using standard telluric spectra obtained in similar conditions of airmass and instrumental set-up of the target observations. This correction has been accomplished with the IRAF\footnote{IRAF is distributed by National Optical Astronomy Observatories, which is operated by the Association of Universities for Research in Astronomy, Inc., under cooperative agreement with the National Science Foundation.} task {\fontfamily{phv}\selectfont telluric}, using spectra of telluric standards from which photospheric lines were removed using a multigaussian fitting. The correction is very good at all wavelengths, with only two regions in the NIR arm ($\lambda\lambda$ 1330-1550 nm, $\lambda\lambda$ 1780-2080 nm) where the telluric absorption bands saturate. More detail about the reduction will be reported in \citet{Alcala13}.

\begin{landscape}
\begin{table}[!]
\caption{\label{known_parameters_table}Known parameters from the literature.}
\begin{tabular}{llcccccccccccccc}
\hline\hline
Name & Other names & RA (J2000) & DEC (J2000) & Region & D & SpT & U & B & V & R & I & J & H & K & Ref \\
\hline
TWA9A & CD $-$36 7429A & 11 48 24.22 & $-$37 28 49.15 & TW Hya &       68 & K5 & ... & 12.52 & 11.26 & ... & 6.94 & 8.68 & 8.03 & 7.85 & 1,2,3 \\
SO879 & ... & 05 39 05.42 & $-$02 32 30.34 & $\sigma$ Ori &      360 & K5 & 17.09 & 14.70 & 14.44 & 13.53 & 12.83 & 11.55 & 10.86 & 10.67 & 3,4,5,6,7,8,9 \\
TWA6 & Tyc 7183 1477 1 & 10 18 28.70 & $-$31 50 02.85 & TW Hya &       51 & M0 & ... & 13.39 & 11.81 & ... & 9.94 & 8.87 & 8.18 & 8.04 & 1,3,10 \\
TWA25 & Tyc 7760 283 1 & 12 15 30.71 & $-$39 48 42.56 & TW Hya &       54 & M0 & ... & 12.85 & 11.44 & ... & 9.50 & 8.17 & 7.50 & 7.31 & 1,3,11,12 \\
TWA14 & UCAC2 12427553 & 11 13 26.22 & $-$45 23 42.74 & TW Hya &       96 & M0.5 & ... & ... & 13.80 & ... & 10.95 & 9.15 & 8.73 & 8.50 & 2,3,13 \\
TWA13B & RX J1121.3$-$3447S & 11 21 17.24 & $-$34 46 45.5 & TW Hya &       59 & M1 & ... & 12.88 & 11.46 & ... & 9.57 & 8.43 & 7.73 & 7.49 & 1,3,11,12 \\
TWA13A & RX J1121.3$-$3447N & 11 21 17.4 & $-$34 46 50 & TW Hya &       59 & M1 & ... & 13.43 & 11.96 & ... & 9.88 & 8.43 & 7.68 & 7.46 & 1,3,11,12 \\
TWA2A & CD $-$29 8887A & 11 09 13.81 & $-$30 01 39.8 & TW Hya &       47 & M2 & ... & 12.55 & 11.07 & ... & 9.20 & 7.63 & 6.93 & 6.71 & 1,2,3 \\
Sz122 & ... & 16 10 16.42 & $-$39 08 05.07 & Lup III &      200 & M2 & ... & 15.33 & 13.73 & 13.28 & 12.12 & 10.89 & 10.12 & 9.93 & 3,14,15,16 \\
TWA9B & CD $-$36 7429B & 11 48 23.73 & $-$37 28 48.5 & TW Hya &       68 & M1 & ... & 15.43 & 14.00 & ... & 11.45 & 9.98 & 9.38 & 9.15 & 1,2,3 \\
TWA15B & ... & 12 34 20.47 & $-$48 15 19.5 & TW Hya &      111 & M2 & ... & 13.30 & 12.20 & 13.41 & 11.80 & 10.49 & 9.83 & 9.56 & 3,17,18,19 \\
TWA7 & Tyc 7190 2111 1  & 10 42 30.06 & $-$33 40 16.62 & TW Hya &       28 & M2 & ... & 12.21 & 10.91 & ... & 9.10 & 7.79 & 7.12 & 6.90 & 1,3,10 \\
 TWA15A & ... &12 34 20.65 & $-$48 15 13.5 & TW Hya &      111 & M1.5 & ... & 13.30 & 12.20 & 13.51 & 11.90 & 10.56 & 9.93 & 9.67 & 3,18,19 \\
Sz121 & ... & 16 10 12.19 & $-$39 21 18.11 & Lup III &      200 & M3 & ... & 15.70 & 14.06 & 13.88 & 11.84 & 10.08 & 9.31 & 9.03 & 3,15,16 \\
Sz94 & ... & 16 07 49.59 & $-$39 04 28.79 & Lup III &      200 & M4 & ... & 16.31 & 16.00 & 14.76 & 12.90 & 11.45 & 10.81 & 10.56 & 3,15,16 \\
SO797 & ... & 05 38 54.92 & $-$02 28 58.35 & $\sigma$ Ori &      360 & M4 & 21.05 & 19.84 & 18.61 & 17.26 & 15.50 & 13.80 & 13.20 & 12.87 & 3,5,20,21 \\
SO641 & ... & 05 38 38.58 & $-$02 41 55.86 & $\sigma$ Ori &      360 & M5 & 21.65 & ... & ... & 18.28 & 16.36 & 14.56 & 13.97 & 13.65 & 3,5,22,23 \\
Par$-$Lup3$-$2 & ... & 16 08 35.78 & $-$39 03 47.91 & Lup III &      200 & M6 & ... & 16.88 & 15.49 & 15.02 & 13.09 & 11.24 & 10.73 & 10.34 & 3,15,24 \\
SO925 & ... & 05 39 11.39 & $-$02 33 32.78 & $\sigma$ Ori &      360 & M5.5 & 22.57 & ... & ... & ... & 16.54 & 14.45 & 13.93 & 13.57 &3,5,9,22 \\
SO999 & ... & 05 39 20.23 & $-$02 38 25.87 & $\sigma$ Ori &      360 & M5.5 & 21.41 & 21.12 & 18.97 & 17.57 & 15.56 & 13.61 & 13.04 & 12.78 & 3,5,21,22 \\
Sz107 & ... & 16 08 41.79 & $-$39 01 37.02 & Lup III &      200 & M5.5 & ... & 17.31 & 16.06 & 15.42 & 13.20 & 11.25 & 10.62 & 10.31 & 3,15,16,25 \\
Par$-$Lup3$-$1 & ... & 16 08 16.03 & $-$39 03 04.29 & Lup III &      200 & M7.5 & ... & 17.74 & 19.92 & 18.15 & 15.28 & 12.52 & 11.75 & 11.26 & 3,15,24 \\
TWA26 & 2M J1139511$-$315921 & 11 39 51.14 & $-$31 59 21.50 & TW Hya &       42 & M9 & ... & 20.10 & ... & 18.10 & 15.83 & 12.69 & 12.00 & 11.50 & 3,26,27 \\
TWA29 & DENIS$-$P J124514.1$-$442907 & 12 45 14.16 & $-$44 29 07.7 & TW Hya &       79 & M9.5 & ... & ... & ... & ... & 18.00 & 14.52 & 13.80 & 13.37 & 28,29 \\
\hline
\end{tabular}
\tablebib{(1)~\citet{Torres06}; (2)~\citet{Barrado06}; (3)~\citet{Cutri03}; (4)~\citet{Caball10}; (5)~\citet{Rigliaco11a}; (6)~\citet{Morrison11}; (7)~\citet{Hernandez07}; (8)~\citet{Sacco08}; (9)~\citet{Caball08}; (10)~\citet{Hog00}; (11)~\citet{Zuckerman04}; (12)~\citet{Messina10}; (13)~\citet{Riaz06}; (14)~\citet{Gomez03}; (15)~\citet{Merin08}; (16)~\citet{Cieza07}; (17)~\citet{Shkolnik11}; (18)~\citet{Samus03}; (19)~\citet{Zuckerman01}; (20)~\citet{Oliveira06}; (21)~\citet{Sherry04}; (22)~\citet{Rigliaco12}; (23)~\citet{Burningham05}; (24)~\citet{Comeron03}; (25)~\citet{Hughes94}; (26)~\citet{Reid08}; (27)~\citet{Denis}; (28)~\citet{Kirkpatrick08}; (29)~\citet{Looper07}; }

\tablefoot{Distances to TW Hya objects are obtained by \citet{Weinberger12}, by \citet{Torres08}, and by \citet{Mamajek05}, to $\sigma$Ori by \citet{Brown}, and to Lupus~III by \citet{Lupus}. }
\end{table}
\end{landscape}

The fully reduced, flux- and wavelength-calibrated spectra are available on the ESO Archive\footnote{http://www.eso.org/sci/observing/phase3/data\_releases.html}.

\begin{table}	
\caption{\label{observations_table}Details of the observations. }
\centering
\begin{tabular}{lccccc}
\hline\hline
Name & \multicolumn{3}{c}{SLITS } &       t$_{\rm exp}$ & Observation \\
\hbox{}  &   UVB  &  VIS  &  NIR  & \hbox{}   &   Date \\
\hline
TWA9A	&0.5\arcsec &  0.4\arcsec  & 0.4\arcsec  &   150s  &    16-17 May 2010\\
SO879		& 1.0\arcsec & 0.9\arcsec & 0.9\arcsec & 3600s  & 11-12 Jan 2011\\
TWA6  &  0.5\arcsec  &   0.4\arcsec  &   0.4\arcsec  &     100s  & 12 Jan 2011\\
TWA25	& 0.5\arcsec  & 0.4\arcsec &  0.4\arcsec   &  120s    &   16-17 May 2010 \\
TWA14 &  0.5\arcsec &  0.4\arcsec &  0.4\arcsec &   400s &  12 Jan 2011\\
TWA13B  	& 0.5\arcsec  & 0.4\arcsec &  0.4\arcsec  &   150s  &    16-17 May 2010 \\
TWA13A 	&0.5\arcsec  & 0.4\arcsec  & 0.4\arcsec  &   150s   &   16-17 May 2010 \\
TWA2A	& 0.5\arcsec &  0.4\arcsec  & 0.4\arcsec  &   100s  &   16-17 May 2010 \\
Sz122  & 1.0\arcsec &  0.9\arcsec &  0.9\arcsec &   600s &  18 Apr 2012\\
TWA9B	&0.5\arcsec  & 0.4\arcsec &  0.4\arcsec   & 800s  &    16-17 May 2010 \\
TWA15B		& 0.5\arcsec  & 0.4\arcsec  & 0.4\arcsec  &  600s  &    16-17 May 2010 \\
TWA7 &   0.5\arcsec  &  0.4\arcsec  &  0.4\arcsec  &    100s  & 12 Jan 2011\\
TWA15A		&0.5\arcsec &  0.4\arcsec  & 0.4\arcsec   & 600s   &   16-17 May 2010 \\
Sz121  & 1.0\arcsec &  0.9\arcsec &  0.9\arcsec &   500s  & 18 Apr 2012 \\
Sz94	 &  1.0\arcsec  & 0.9\arcsec  & 0.9\arcsec   & 600s     & 16-17 May 2010 \\
SO797	& 1.0\arcsec  & 0.9\arcsec  & 0.9\arcsec  & 2400s   &   23-24 Dec 2009 \\
SO641	& 1.0\arcsec  & 0.9\arcsec  & 0.9\arcsec  &  3600s  &    23-24 Dec 2009 \\
Par$-$Lup3$-$2	&  1.0\arcsec &  0.9\arcsec &  0.9\arcsec  &  1200s   &   16-17 May 2010 \\
SO925	&  1.0\arcsec  & 0.9\arcsec  & 0.9\arcsec  &  3600s  &    21-22 Dec 2009 \\
SO999		&  1.0\arcsec  & 0.9\arcsec  & 0.9\arcsec  & 2400s  &    24-25 Dec 2009\\
Sz107		& 1.0\arcsec &  0.9\arcsec  & 0.9\arcsec &   600s   &   22 Apr 2011 \\
Par$-$Lup3$-$1	&  1.0\arcsec  & 0.9\arcsec  & 0.9\arcsec  &  600s  &    16-17 May 2010 \\
TWA26		&1.0\arcsec  & 0.9\arcsec &  0.9\arcsec &   3600s  &    22 Mar 2010 \\
TWA29	& 1.6\arcsec  & 1.5\arcsec &  1.5\arcsec &   3600s   &   22 Mar 2010\\
\hline
\end{tabular}

\end{table}


\section{Spectral type classification}
\label{sec::source_class}

A careful SpT classification of the sample is important in order to provide correct templates for accretion estimates of Class II YSOs. Moreover, the procedure used to derive the SpT of Class II and Class III YSOs should be as homogeneous as possible. In this section we describe two different methods to derive SpT for these objects. Firstly, we use the depth of various molecular bands in the VIS part of the spectrum. Then, we describe the second method, which consists of using spectral indices in the VIS and in the NIR part of the spectrum. These provide us a reliable, fast and reddening free method to determine SpT for large samples of YSOs. 

\subsection{Spectral typing from depth of molecular bands}
\label{subsec::source_class_spt}
For the SpT classification of the objects, we use the analysis of the depth of several molecular bands in the spectral region between 580 nm and 900 nm \citep{Luhman04, Allen95, Henry94}. This region includes various TiO ($\lambda\lambda$ 584.7-605.8, 608-639, 655.1-685.2, 705.3-727, 765-785, 820.6-856.9, 885.9-895 nm), VO ($\lambda\lambda$ 735-755, 785-795, 850-865 nm) and CaH ($\lambda\lambda$ 675-705 nm) absorption bands, and a few photospheric lines (the CaII IR triplet at $\lambda\lambda$ 849.8, 854.2, 866.2 nm, the NaI doublet at $\lambda$ 589.0 and 589.6 nm, the CaI at $\lambda$ 616.2 nm, a blend of several lines of BaII, FeI and CaI at $\lambda$ 649.7 nm, the MgI at $\lambda$ 880.7 nm, and the NaI and KI doublets at $\lambda\lambda$ 818.3 nm and 819.5 and $\lambda\lambda$ 766.5 nm and 769.9, respectively. 

\begin{figure*}[!]
   \centering
   \includegraphics[width=\textwidth]{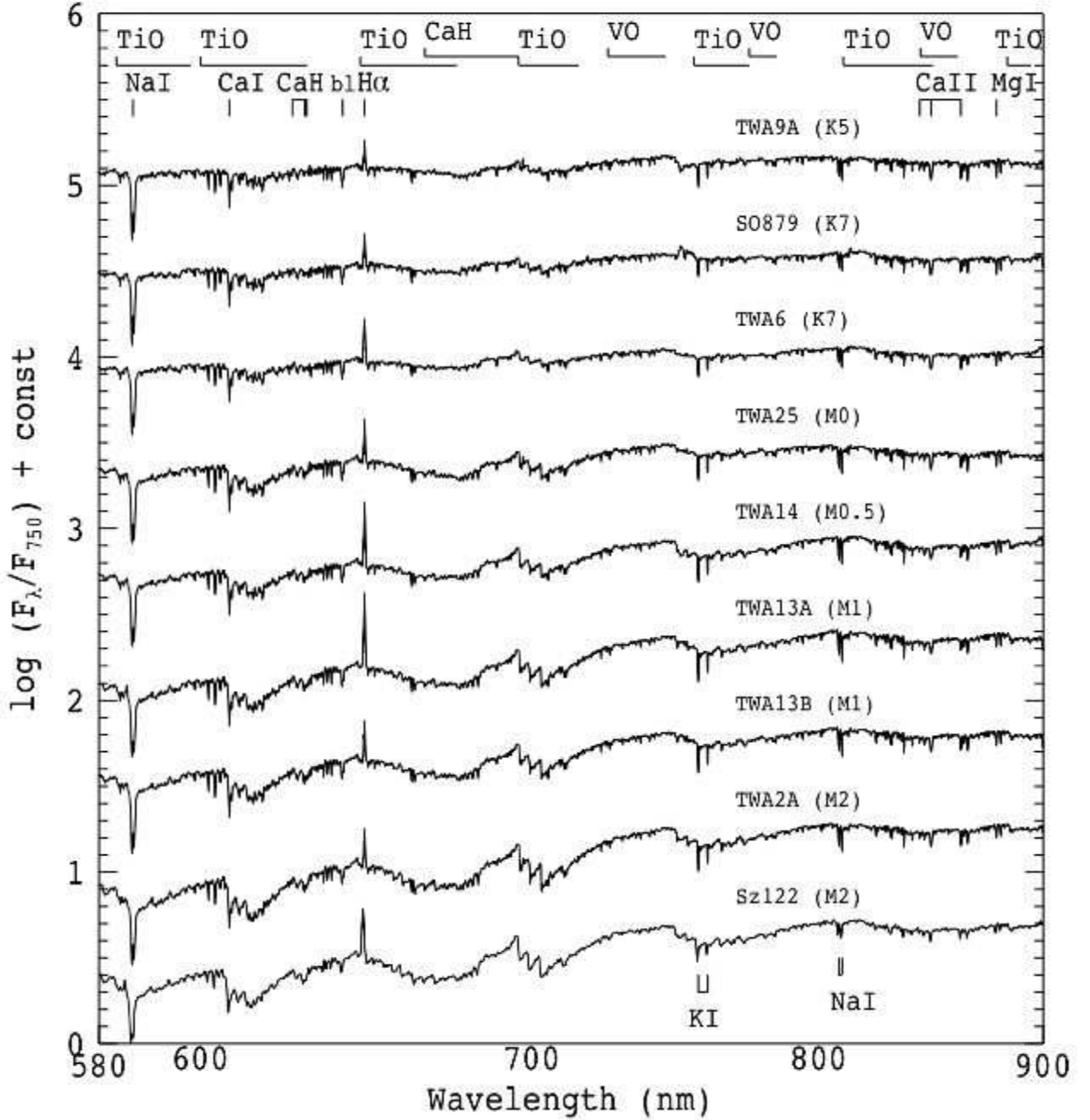}	 
      \caption{Spectra of Class III YSOs with SpT earlier than M3 in the wavelength region where the spectral classification has been carried out (see text for details). All the spectra are normalized at 750 nm and offset in the vertical direction by 0.5 for clarity. The spectra are also smoothed to the resolution of 2500 at 750 nm to make easier the identification of the molecular features.}
         \label{spec}
   \end{figure*}

\begin{figure*}[!]
   \centering
   \includegraphics[width=\textwidth]{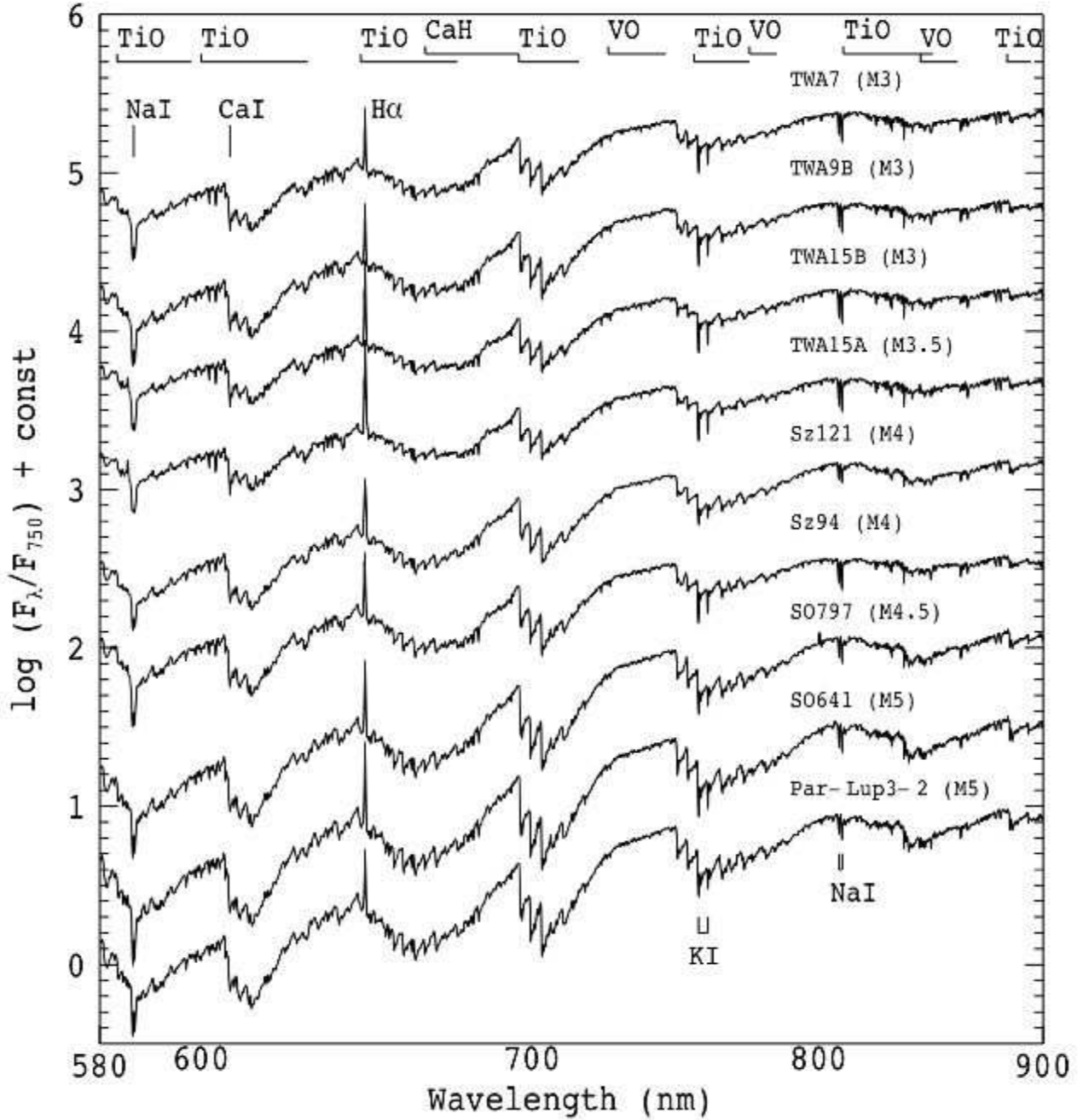}
      \caption{Same as Fig.~\ref{spec}, but for SpTs between M3 and M5. }
	\label{spec2}
   \end{figure*}

\begin{figure*}[!]
   \centering
   \includegraphics[width=\textwidth]{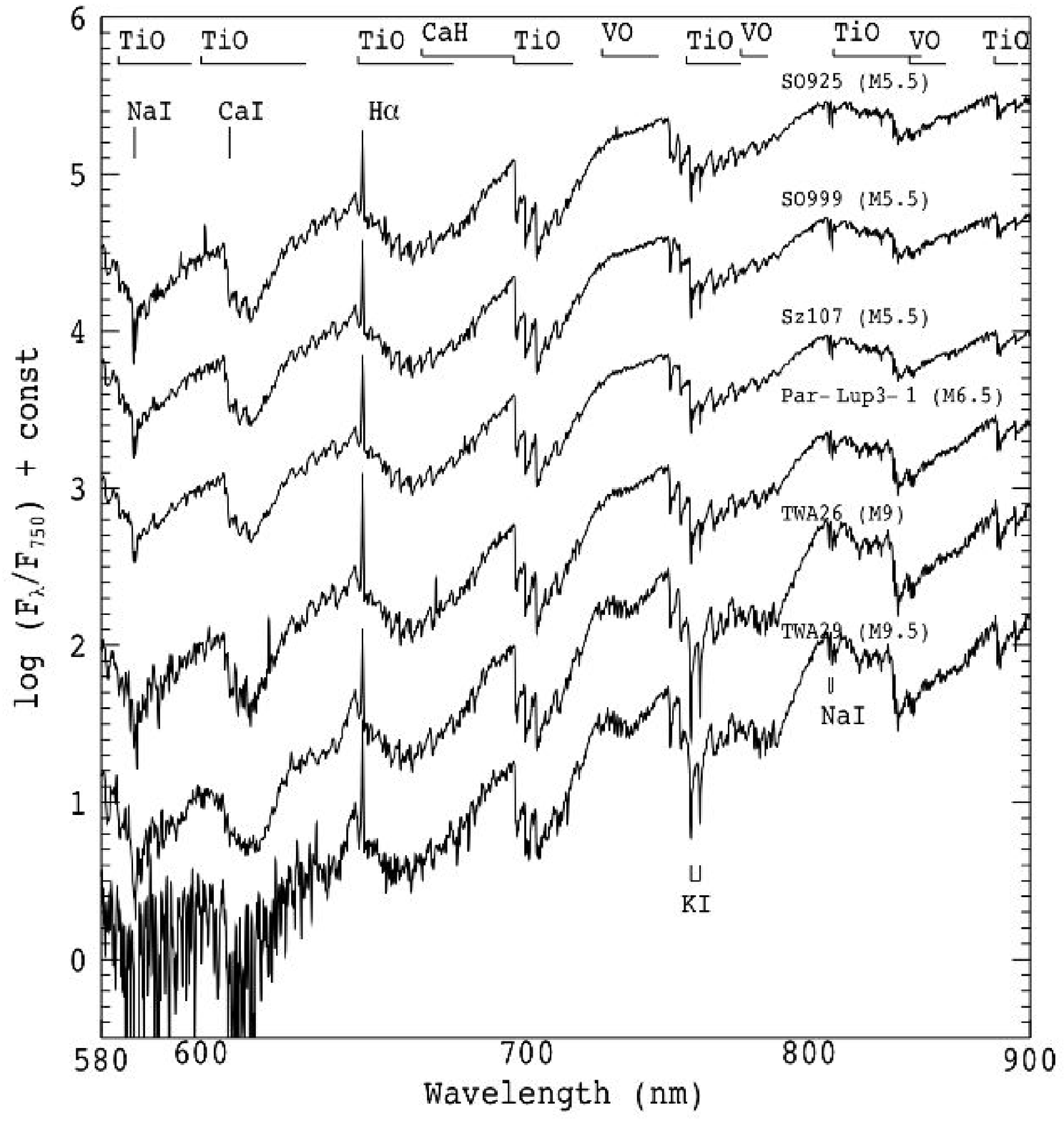}
      \caption{Same as Fig.~\ref{spec}, but for SpTs later than M5. }
	\label{spec3}
   \end{figure*}

In Fig.~\ref{spec}, ~\ref{spec2} and ~\ref{spec3} we show the VIS spectra of the objects in the wavelength range between 580 and 900 nm. All the spectra are normalized at 750 nm and, for the sake of clarity, smoothed to a resolution of R$\sim$2500 at 750 nm and vertically shifted. The depth of the molecular features increases with SpT almost monotonically, and, comparing the spectra of the targets using together different wavelength subranges and different molecular bands, we can robustly assign a SpT to our objects thanks to the differences in the depth of the bands, with uncertainties estimated to be 0.5 subclasses. 

\begin{figure}
   \centering
   \includegraphics[width=0.5\textwidth]{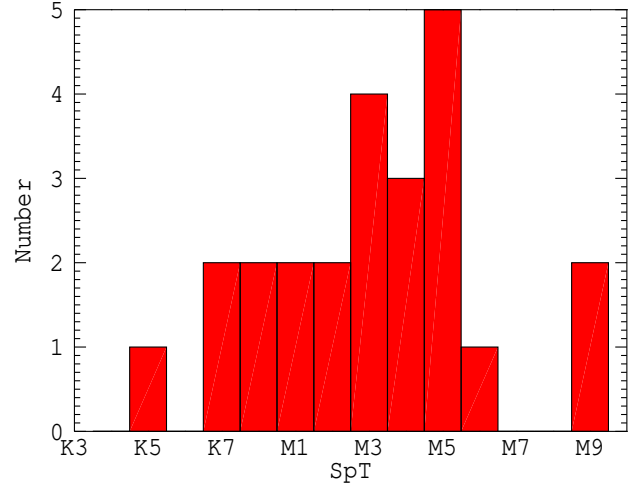}	  
      \caption{Distribution of spectral types of the Class III YSOs discussed in this work. Each bin corresponds to one spectral sub-class.}
         \label{fig::sample}
   \end{figure}

With this spectral typing procedure, we classify all the objects in our sample with SpT earlier than M8. The agreement between the SpTs derived here and those in the literature is good, with only three cases where the difference is 2 spectral subclasses.  We assume the SpT available from the literature for the 2 YSOs with SpT later than M8 \citep{Reid08,Kirkpatrick08}, because the classification by comparison of molecular bands depth with other spectra in the sample is not possible due to the fact that our sample has a gap between M6.5 and M9. We can only confirm that these objects have a SpT later than the other targets in the sample and that their SpT differ by 0.5 subclass. The SpT obtained here are listed in Table~\ref{sample_parameters} and in the first column of Table~\ref{tab::NIR_indices_results}, and they are used for the rest of the analysis. 
The distribution of SpT for the sample is shown in Fig.~\ref{fig::sample}. The range of the M-type is almost entirely covered, providing a good sample for the goals of this paper and a solid library of templates that can be used for \lacc \ estimates of Class II YSOs.

In Appendix~\ref{app::spectra} the NIR and UVB spectra of all the sources are shown. Also in these cases a trend with SpT can be seen.

\begin{table*}[!]	
\caption{\label{sample_parameters}Stellar parameters derived for the objects in our sample}
\centering
\begin{tabular}{lccccc}
\hline\hline
Name & SpT & T$_{\rm eff}$ [K] & log($L_*/L_\odot$) & $M_*$[$M_\odot$] & $<$log(L$_{acc,noise}$/L$_\odot$)$>$ \\
\hline
TWA9A & K5 &     4350 & $-$0.61 & 0.81 & $-$3.38 \\
SO879 & K7 &     4060 & $-$0.29 & 1.07 & $-$2.88 \\
TWA6 & K7 &     4060 & $-$0.96 & 0.66 & $-$3.37 \\
TWA25 & M0 &     3850 & $-$0.61 & 0.84 & $-$3.04 \\
TWA14 & M0.5 &     3780 & $-$0.83 & 0.73 & $-$3.23 \\
TWA13B & M1 &     3705 & $-$0.70 & 0.68 & $-$3.32 \\
TWA13A & M1 &     3705 & $-$0.61 & 0.70 & $-$2.85 \\
TWA2A & M2 &     3560 & $-$0.48 & 0.55 & $-$3.36 \\
Sz122 & M2 &     3560 & $-$0.60 & 0.54 & $-$2.57 \\
TWA9B & M3 &     3415 & $-$1.17 & 0.37 & $-$3.93 \\
TWA15B & M3 &     3415 & $-$0.96 & 0.37 & $-$3.45 \\
TWA7 & M3 &     3415 & $-$1.14 & 0.37 & $-$3.84 \\
TWA15A & M3.5 &     3340 & $-$0.95 & 0.30 & $-$3.23 \\
Sz121 & M4 &     3270 & $-$0.34 & 0.37 & $-$3.25 \\
Sz94 & M4 &     3270 & $-$0.76 & 0.28 & $-$3.38 \\
SO797 & M4.5 &     3200 & $-$1.26 & 0.19 & $-$4.27 \\
SO641 & M5 &     3125 & $-$1.53 & 0.12 & $-$4.51 \\
Par$-$Lup3$-$2 & M5 &     3125 & $-$0.75 & 0.18 & $-$3.89 \\
SO925 & M5.5 &     3060 & $-$1.59 & 0.10 & $-$4.65 \\
SO999 & M5.5 &     3060 & $-$1.28 & 0.13 & $-$4.30 \\
Sz107 & M5.5 &     3060 & $-$0.79 & 0.16 & $-$3.69 \\
Par$-$Lup3$-$1 & M6.5 &     2935 & $-$1.18 & 0.10 & $-$4.74 \\
TWA26 & M9 &     2400 & $-$2.70 & 0.02 & $-$6.54 \\
TWA29 & M9.5 &     2330 & $-$2.81 & 0.02 & $-$6.64 \\
\hline
\end{tabular}

\tablefoot{The spectral type-T$_{\rm eff}$ relation is adopted from \citet{Luhman03} for M-type objects and from \citet{Kenyon95} for K-type objects.}

\end{table*}

\begin{table*}
\caption{\label{tab::NIR_indices_results}Spectral types obtained using the method based on the spectral indices described in Sect.~\ref{sec::source_class} and in Appendix~\ref{NIR_ind_app}.}
\centering
\begin{tabular}{lcccccccccc}
\hline\hline
Name & SpT\tablefootmark{a} & VIS\_ind\tablefootmark{b} & H$_2$O\_K2 & H$_2$O & sH$_2$O$^J$ & sH$_2$O$^K$ & sH$_2$O$^{H1}$ & I$_J$ & I$_H$ & HP \\
\hline
TWA9A & K5 & ... & ... & ... & ... & ... & ... & ... & ... & ... \\
SO879 & K7 & ... & ... & ... & ... & ... & ... & ... & ... & ... \\
TWA6 & K7 & ... & ... & ... & ... & ... & ... & ... & ... & ... \\
TWA25 & M0 & ... & ... & ... & ... & ... & ... & ... & ... & ... \\
TWA14 & M0.5 & ... & ... & ... & ... & ... & ... & ... & ... & ... \\
TWA13B & M1 & M3.3 &... & ... & ... & M1.0 & M1.2 & ... & M0.5 & ... \\
TWA13A & M1 & M3.3 & ... & ... & ... & M1.0 & M0.7 & ...& M0.7 & ... \\
TWA2A & M2 & M3.3 & M2.0 & ... & M2.6 & M1.5 & M0.1 & M1.4 & M0.5 & ... \\
Sz122 & M2 & M3.4 & M2.2 & ... & M2.6 & M2.7 & M2.4 & M2.5 & M2.0 & ... \\
TWA9B & M3 & M3.7 & M4.0 & ... & M5.1 & M3.1 & M3.4 & M4.4 & M3.4 & ... \\
TWA15B & M3 & M3.8 & M4.5 & ... & M6.7 & M3.8 & M3.3 & M5.0 & M2.1 & ... \\
TWA7 & M3 & M3.8 & M2.6 & ... & M3.5 & M4.3 & M3.3 & M3.3 & M3.7 & ... \\
TWA15A & M3.5 & M3.8 & M3.8 & ... & M7.2 & M4.1 & M4.0 & M5.3 & M2.5 & ... \\
Sz121 & M4 & M4.3 & M3.0 & ... & M3.6 & M1.2 & M5.8 & M2.9 & M2.1 & ... \\
Sz94 & M4 & M3.7 & M4.3 & ... & M4.9 & M3.7 & M3.0 & M4.9 & M4.0 & ... \\
SO797 & M4.5 & M4.7 & M2.3 &...& M2.1 & M6.0 & M5.0 & M3.6 & M5.9 & ... \\
SO641 & M5 & M5.2 & M5.3 & M5.8 & M4.7 & M4.2 & M5.1 & M5.9 & M6.6 & ... \\
Par$-$Lup3$-$2 & M5 & M5.0 & M5.7 & M5.7 & M5.8 & M4.0 & M4.5 & M6.1 & M5.0 & ... \\
SO925 & M5.5 & M5.5 & M7.3 & M6.3 & M3.9 & M1.5 & M5.0 & M6.2 & M7.3 & ... \\
SO999 & M5.5 & M5.4 & M3.6 & M5.8 & M3.9 & M5.7 & M5.4 & M6.1 & M6.5 & ... \\
Sz107 & M5.5 & M5.5 & M5.7 & M5.9 & M6.1 & M6.1 & M5.6 & M6.4 & M4.5 & ... \\
Par$-$Lup3$-$1 & M6.5 & M6.4 & M6.6 & M7.1 & L0.8 & M5.7 & M8.6 & L0.0 & M5.5 & ... \\
TWA26 & M9 & ... & M8.4 & M8.3 & L0.0 & M7.7 & M9.9 & L0.4 & M6.8 & M9.1 \\
TWA29 & M9.5 & ... & M9.4 & M8.7 & M9.3 & M8.2 & L0.8 & L0.2 & M7.2 & M9.5 \\
\hline
\end{tabular}
\tablefoot{\tablefoottext{a}{SpT derived in this work as explained in Sect.~\ref{subsec::source_class_spt}.} \tablefoottext{b}{Results obtained using the spectral indices in the VIS part of the spectrum, as explained in Sect.~\ref{subsec::source_class_ind}.} All the other columns refer to the results obtained using NIR spectral indices, as explained in Appendix~\ref{NIR_ind_app}. SpT are reported only in the range of validity of each index. }

\end{table*}

\subsection{Spectral indices for M3-M8 stars}
\label{subsec::source_class_ind}
Spectral indices provide a fast method to determine SpT for large samples of objects. Here we test some of these indices for M-type objects. \citet{Riddick07} tested and calibrated various spectral indices in the VIS part of the spectrum for pre main sequence (PMS) stars with SpT from M0.5 to M9. In fact, they suggest the use of some reliable spectral indices valid in the range M3-M8. We find that the best SpT classification can be achieved by combining together results obtained using the set of indices that we report in Table~\ref{indices_table}. We proceed as follows. For each object we calculate the SpT with all these indices, and then assign the mean SpT using those results that are in the nominal range of validity of each index. Typical dispersions of the SpT derived with each index are less than half a subclass. We report the final results obtained with these indices in the second column (VIS\_ind) of Table~\ref{tab::NIR_indices_results}. Comparing these results with those derived in the previous section, we report an agreement within one subclass for all the objects in the range M3-M8, as expected. However, there are 4 YSOs classified M1-M2 from the depth of molecular bands that would be classified M3 from the values of the Riddick's indices. This suggests that spectral classification M3 obtained with spectral indices could, in fact, be earlier by more than one subclass.

There are also indices based on features in the NIR part of the spectrum \citep[e.g][]{Testi01,Testi09,Allers07,Rojas-Ayala12}. In Appendix~\ref{NIR_ind_app} we compare the results obtained with these NIR indices with the SpT derived in the previous section, in order to confirm the validity of some of these indices for the classification of M-type YSOs. 

\begin{table*}[!]
\caption{\label{indices_table}Spectral indices from \citet[et reference therein]{Riddick07} adopted in  our analysis for spectral type classification.}
\centering
\begin{tabular}{lccc}
\hline\hline
Index&Range of validity&Numerator [nm]&Denominator [nm] \\
\hline
VO 7445          & M5-M8     & 0.5625 (735.0-740.0) + 0.4375 (751.0$-$756.0) & 742.0$-$747.0 \\
VO 2          & M3-M8 &792.0$-$796.0& 813.0$-$815.0\\
c81 & M2.5-M8     &811.5$-$816.5&(786.5$-$791.5)+(849.0$-$854.0)\\
R1         & M2.5-M8      &802.5$-$813.0 &801.5$-$802.5\\
R2         & M3-M8      &814.5$-$846.0&846.0$-$847.0\\
R3         & M2.5-M8     &(802.5$-$813.0)+(841.5$-$846.0)&(801.5$-$802.5)+(846.0$-$847.0)\\
TiO 8465        & M3-M8      &840.5$-$842.5&845.5$-$847.5 \\
PC3        & M3-M8      &823.5$-$826.5 &754.0$-$758.0\\
\hline
\end{tabular}
\end{table*}


\section{Stellar parameters}
\label{sec::star_param}
To estimate the mass ($M_*$), radius ($R_*$) and age for each target we compare the photospheric parameters ($L_*$,T$_{\rm eff}$) with the theoretical predictions of the \citet{Baraffe98} PMS evolutionary tracks. We derive the T$_{\rm eff}$ of each star from its SpT using the \citet{Luhman03} SpT-T$_{\rm eff}$ scale, while the procedure to estimate $L_*$ is described in the next paragraph. We then use ($L_*$,T$_{\rm eff}$) to place the stars on the HR diagram and estimate ($M_*$, age) interpolating the theoretical evolutionary tracks. These parameters are reported in Table~\ref{sample_parameters}. Finally, we obtain $R_*$ from $L_*$ and T$_{\rm eff}$.

\subsection{Stellar luminosity}
\label{lum_sec}
\begin{figure}
   \centering
   \includegraphics[width=0.5\textwidth]{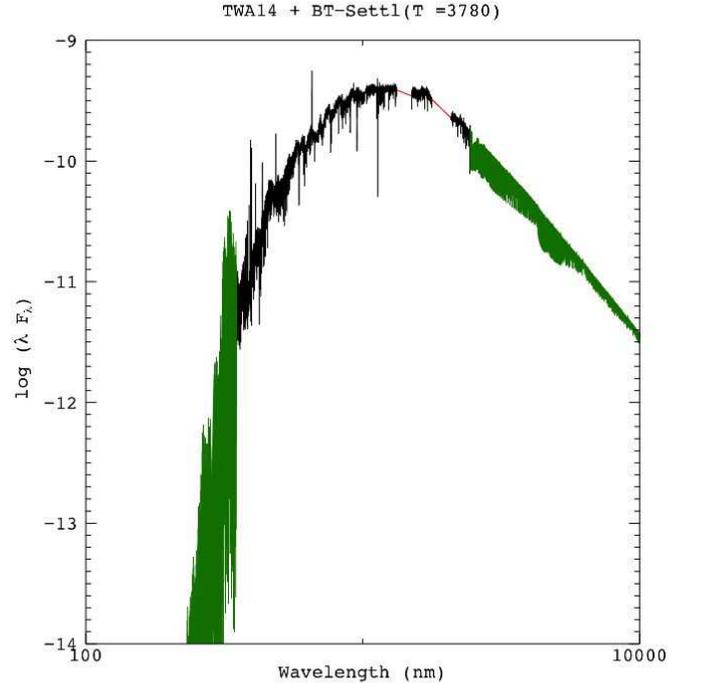} 
      \caption{Example of the combination of a flux-calibrated and telluric removed X-Shooter spectrum (black) with model spectra with the same T$_{\rm eff}$ (green), which is normalized at the red and blue edges of the X-Shooter spectrum and shown only outside the X-Shooter range. The telluric bands in the NIR are replaced with a linear interpolation (red). In this figure we show the example of TWA14.}
         \label{lum_example_fig}
   \end{figure}

Given the large wavelength coverage ($\sim$300-2500 nm) of the X-Shooter spectra, for objects with 2300 $<$ T$_{\rm eff}<$ 4400 K only a small percentage of the stellar flux ($\lesssim$ 10-30\%) arises from spectral regions outside the X-Shooter spectral range. We use, therefore, the following procedure to estimate the total flux of our objects: first, we integrate the whole X-Shooter spectrum from 350 nm to 2450 nm, excluding the last 50 nm of spectra on each side, that are very noisy, and the regions in the NIR between the $J$, $H$ and $K$-bands ($\lambda\lambda$ 1330-1550 nm, $\lambda\lambda$ 1780-2080 nm, see Sect.~\ref{observ_sec}), where we linearly interpolate across the telluric absorption regions. We then use the BT-Settl synthetic spectra from \citet{Allard11} with the same T$_{\rm eff}$ as our targets (see Table~\ref{sample_parameters}), assuming log$g$=4.0, typical of low-mass YSOs, and normalized with our spectra at 350 and 2450 nm to estimate the contribution to the total flux emitted outside the observed range. The match of the normalization factors at the two ends is in all cases very good. We show in Fig.~\ref{lum_example_fig} an example of this procedure. 

The main source of uncertainty in $L_*$ comes from the uncertainty in the spectroscopic flux. For the objects observed in excellent weather conditions this matches the fotometric fluxes to better than a factor $\lesssim$1.5. It is then reasonable to assume such an uncertainty for all the objects in our sample, after normalizing the spectroscopic flux to the photometry (see Sect.~\ref{sec::data_red}). This would lead to an uncertainty of less than 0.2 dex in log$L_*$. 

To convert the bolometric fluxes obtained in this way in $L_*$ we adopt these distances: we assume that YSOs in the $\sigma$Ori region have a distance of 360 pc \citep{Brown}, those in TW Hya the distances listed in \citet{Weinberger12}, in \citet{Torres08}, and in \citet{Mamajek05}, and those in Lupus III of 200 pc \citep{Lupus}, as reported in Table~\ref{known_parameters_table}. The derived stellar luminosities are listed in Table~\ref{sample_parameters}. These values are comparable with those obtained with photometric data in the literature, with a typical difference $\lesssim$ 0.2 dex. Hence, they are consistent with our determinations, within the errors.

\subsection{Stellar mass and age}
\begin{figure}
   \centering
   \includegraphics[width=0.5\textwidth]{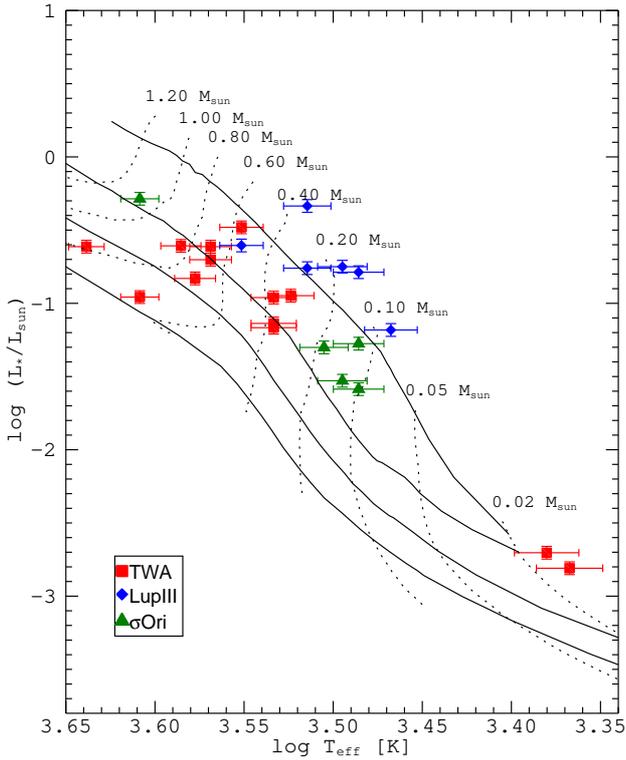} 
      \caption{Hertzsprung-Russell Diagram of the Class III YSOs of this work. Data points are compared with the evolutionary tracks (dotted lines) by \citet{Baraffe98}. Isochrones (solid lines) correspond to 2, 10, 30 and 100 Myr.}
         \label{HRD_plot}
   \end{figure}

In Fig.~\ref{HRD_plot} we show the Hertzsprung-Russell diagram (HRD) of our PMS stars, built using $T_{\rm eff}$ as reported in Table~\ref{sample_parameters} and $L_*$ derived in Sec~\ref{lum_sec}.  We assign $M_*$ and age to our PMS stars by interpolating evolutionary tracks from \citet{Baraffe98} in the HRD.  The resulting $M_*$ are reported in Table~\ref{sample_parameters}. For four objects (Sz107, Sz121, TWA26, and TWA29) the position in the HRD implies an age $<$1 Myr, where theoretical models are known to be very uncertain, and in fact are typically not tabulated \citep{Baraffe98}. We estimate the mass of these objects extrapolating from the closest tabulated points, but we warn the reader that the values are affected by high uncertainty.

Our YSOs are distributed along different isochrones; Lupus YSOs appear to be younger than the others (age$\lesssim$2 Myr), while $\sigma$Ori YSOs are distributed in isochronal ages in the range 2$\lesssim$age$\lesssim$10 Myr; finally, TW Hya targets appear generally close to the 10 Myr isochrone. This is in general agreement with what found in the literature; indeed, Lupus has an estimated age of $\sim$1-1.5 Myr \citep{Hughes94, Lupus}, while the $\sigma$Ori region is usually considered to be slightly older ($\sim$3 Myr in average), and ranging from $\lesssim$1 Myr to several Myr \citep{Zapatero-Osorio02,Oliveira04}.  For the TW Hya association, the age estimates are $\gtrsim$10 Myr \citep{Mamajek05,Barrado06,Weinberger12}.

\section{Line classification}
\label{sec::lines}
The spectra of our objects are characterized by the presence of photospheric absorption lines that depend on the SpT and in some cases on the age. In order to assess the PMS status of the objects in our sample, we check that the lithium absorption feature at $\lambda$ 670.8 nm, which is related to the age of the YSOs \citep[e.g.][]{Mentuch08}, is detected in all but one (Sz94) of the objects. We discuss in more detail the implications of the non-detection in Sz94 in Appendix~\ref{single_sources}, and we explain why this object could be anyway considered in our analysis as YSO. The values of the lithium equivalent width (EW$_{\rm LiI}$) for the other objects in the sample are $\sim$ 0.5 \AA. A detailed analysis of this line and the other photospheric absorption lines of the objects in our sample will be carried out by \citet{Stelzer13b}. 

In addition to these, we detect many emission lines, typically H, He and Ca lines, that originate in the chromosphere of these stars. In this work we concentrate on the emission lines characterization, as we are interested in the chromospheric activity.

\subsection{Emission lines identification}
\label{subsec::em_lines}
In order to understand the contribution of the chromospheric emission to the estimate of \lacc \ through the luminosity of accretion-related emission lines, we first identify in our spectra the lines typically related to accretion processes in Class II YSOs. Here, we describe which lines we detect and report their fluxes and equivalent widths in Table~\ref{line_fluxes_table} and~\ref{line_fluxes_table_others}.

\begin{figure}
   \centering
   \includegraphics[width=0.5\textwidth]{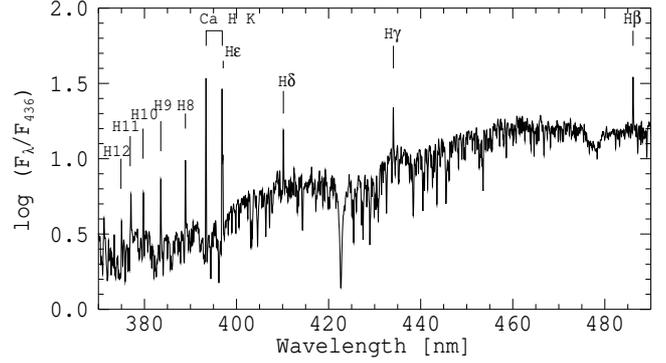} 
      \caption{Portion of the spectrum, showing emission in all Balmer lines from H$\beta$ up to H12, as well as the CaII H and K lines, of the YSO TWA13B. The spectrum has been smoothed to a resolution R = 3750 at 375 nm.}
         \label{Balmer_decrement}
   \end{figure}

The most common line detected in YSOs is the H$\alpha$ line at 656.28 nm, that is present in the spectra of all our objects. Emission in this line has been used as a proxy for YSO identification and has been related to accretion processes \citep[e.g.][]{Muzerolle98,Natta04}. This line is also generated in chromospherically active YSOs \citep[e.g.][]{WB03}. Similarly, the other hydrogen recombination lines of the Balmer series are easily detected in almost all of our Class III objects up to the H12 line ($\lambda$ 374.9 nm). It is not easy, nevertheless, to determine the continuum around Balmer lines beyond the H9 line ($\lambda$ 383.5 nm), and the H$\epsilon$ line ($\lambda$ 397 nm) is blended with the CaII-K line. An example of a portion of the spectrum from H$\beta$ to H12 is shown in Fig.~\ref{Balmer_decrement}. 

The hydrogen recombination lines of the Paschen and Brackett series, in particular the Pa$\beta$ ($\lambda$ 1281.8 nm) and Br$\gamma$ ($\lambda$ 2166 nm) lines, have been shown to be related to accretion by \citet{Muzerolle98}. These lines have subsequently been used to survey star forming regions with high extinction \citep{Natta04,Natta06} in order to obtain accretion rate estimates for very low mass objects. We do not detect any of these lines in our Class III spectra, confirming that chromospheric activity is not normally detectable with these lines.

The Calcium II emission lines at $\lambda\lambda$ 393.4, 396.9 nm (Ca HK) and at $\lambda\lambda$ 849.8, 854.2, 866.2 nm (Ca IRT) are related to accretion processes \citep[e.g.][]{Mohanty05, Herczeg08, Rigliaco12}, but also to chromospheric activity \citep[e.g.][]{Montes98}. The CaII H and K lines are detected in 90\% of our objects. The CaII IRT lines are detected in 11 out of 13 objects with SpT earlier than M4. These emission lines appear as a reversal in the core of the photospheric absorption lines. For all 11 objects with SpT M4 or later, the CaII IRT lines are not detected.

The HeI line at $\lambda$ 587.6 nm is also known to be associated with accretion processes \citep{Muzerolle98,Herczeg08}, but in Class III YSOs is known to be of chromospheric origin \citep[e.g.][]{Edwards06}. The line is indeed detected in 22 (92\%) objects. Other HeI lines at $\lambda\lambda$ 667.8, 706.5, and 1083 nm are usually associated with accretion processes \citep{Muzerolle98,Herczeg08,Edwards06}; we detect only in Sz122 the HeI lines at $\lambda\lambda$ 667.8, 706.5 nm, while we detect in 8 (33\%) of the objects the HeI line at $\lambda$ 1083 nm.

Finally, there is no trace of forbidden emission lines in any of our X-Shooter spectra, consistent with the expected absence of circumstellar material in Class III YSOs.

\subsection{H$\alpha$ equivalent width and 10\% width}
\begin{figure}
   \centering
   \includegraphics[width=0.5\textwidth]{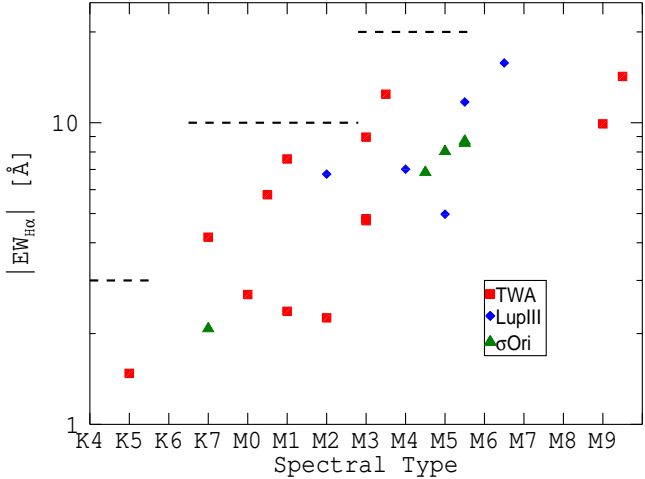} 
      \caption{H$\alpha$ equivalent width as a function of spectral type. The dashed lines represent the boundary between accretors and non accretors proposed by \citet{WB03} for different SpT.}
         \label{Ha_EW}
   \end{figure}

A commonly used estimator for the activity in PMS stars is the EW of the H$\alpha$ line \citep[e.g.][]{WB03}. This is useful especially when dealing with spectra that are not flux-calibrated or with narrow-band photometric data. The absolute values of this quantity as a function of the SpT of the objects are plotted in Fig.~\ref{Ha_EW} and the values are reported in Table~\ref{line_fluxes_table}. We observe a well-known dependence of EW$_{H\alpha}$ with SpT that is due to decreasing continuum flux for cooler atmospheres. With respect to the threshold to distinguish between accreting and non-accreting YSOs proposed by \citet{WB03}, all our targets satisfy the criteria of \citet{WB03} for being non-accretors.

\begin{figure}
   \centering
   \includegraphics[width=0.5\textwidth]{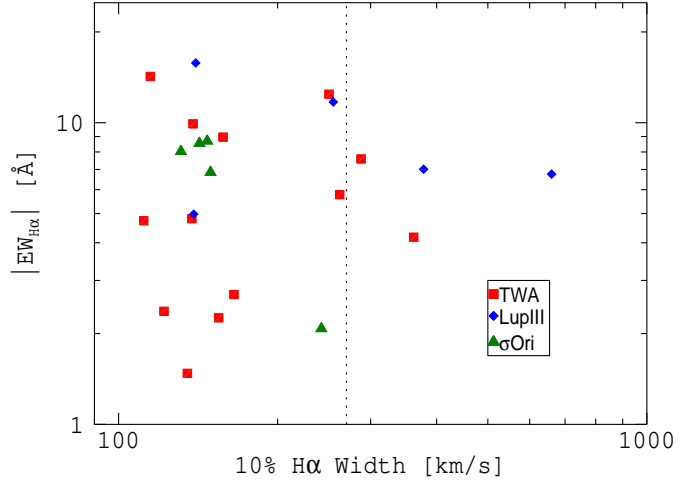} 
      \caption{H$\alpha$ equivalent width as a function of the 10\% H$\alpha$ width. The vertical dashed line rerpesents the \citet{WB03} criterion for the boundary between accretors and non-accretors. The objects with 10\% H$\alpha$ width bigger than  270 km/s are, from right to left: Sz122, Sz121, TWA6 and TWA13A.}
         \label{Ha_EW_vs_10width}
   \end{figure}

Another diagnostic to distinguish between accreting and non-accreting YSOs is the full width of the H$\alpha$ line at 10\% of the line peak \citep{WB03}. This diagnostic has been shown to be correlated with \macc, but with a large dispersion \citep{Natta04}.  Fig.~\ref{Ha_EW_vs_10width} shows the EW$_{H\alpha}$ absolute values versus the 10\% H$\alpha$ width. We see that for most of our objects the 10\% H$\alpha$ width is in the $\sim$100-270 km/s range, and for only three objects this value is significantly above the threshold suggested by \citet{WB03} of 270 km/s (TWA6, Sz122, and Sz121). In Appendix~\ref{single_sources} we discuss about these objects, and we explain why TWA6 and Sz121 can be considered in our analysis, while Sz122 should be excluded because it is probably an unresolved binary. This is probably due to high values of $v$sin$i$ for these objects, which broaden the line profile or to unresolved binarity. For BDs, the threshold to distinguish accretors from non-accretors is at 10\% H$\alpha$ width of 200 km/s \citep{Jayawardhana03}. This is satisfied for all the BDs in the sample. The values of 10\% H$\alpha$ width are listed in Table~\ref{line_fluxes_table}. 

\subsection{Line luminosity}
\label{subsec::line_lum}
We measure the flux of each line by estimating the continuum in the proximity of the line with the IDL astrolib outlier-resistant mean task {\fontfamily{phv}\selectfont resistant\_mean}. We then subtract the continuum from the observed flux and calculate the integral, checking that the whole line, including the wings, is included in the computation. To compute $L_{\rm line}$ we adopt the distances reported in Table~\ref{sample_parameters}. 

For the CaII IRT lines, where the emission appears in the core of the absorption feature, we subtract the continuum from our spectrum following the prescription given by \citet{Soderblom93}. Using a BT-Settl synthetic spectrum \citep{Allard11} of the same $T_{\rm eff}$ smoothed at the same resolution of our observed spectrum, we obtain an estimate of the line absorption feature that is then subtracted in order to isolate the emission core of the line; finally, we integrate over the continuum subtracted spectrum. We report in Tables~\ref{line_fluxes_table} and~\ref{line_fluxes_table_others} the values obtained for the fluxes and the line EWs.

  \begin{figure*}
   \centering
  \includegraphics[width=\textwidth]{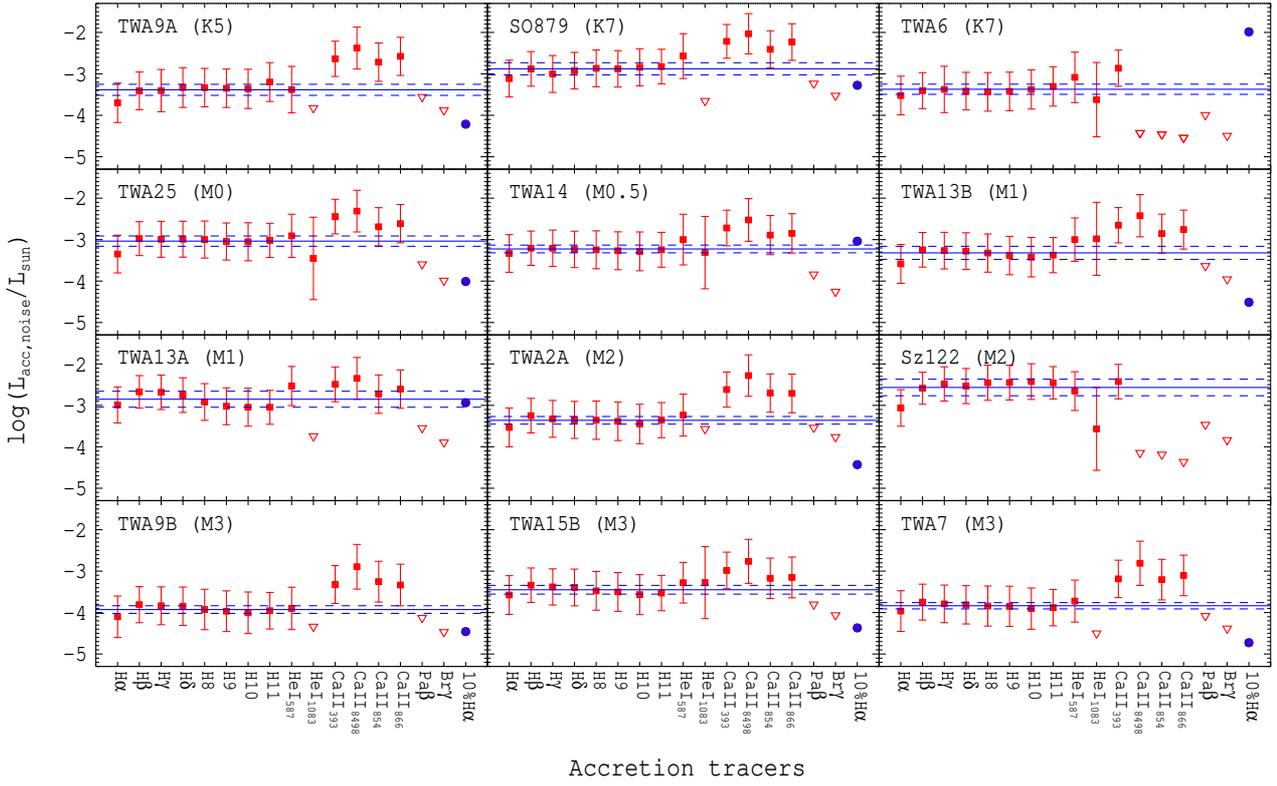} 
      \caption{$\log(L_{\rm acc,noise}/L_\odot)$ obtained using different accretion tracers and the relations between $L_{\rm line}$ and $L_{\rm acc}$ from \citet{Alcala13}. The mean values obtained using the Balmer and HeI$_{\lambda587.6}$ lines are shown with the blue solid lines and the 1$\sigma$ dispersion is reported with the blue dashed lines. Upper limits are reported with red empty triangles. The 10\% H$\alpha$ width is reported with a blue filled circle. }
         \label{Lacc_all_sources1}
   \end{figure*}

  \begin{figure*}
   \centering
 \includegraphics[width=\textwidth]{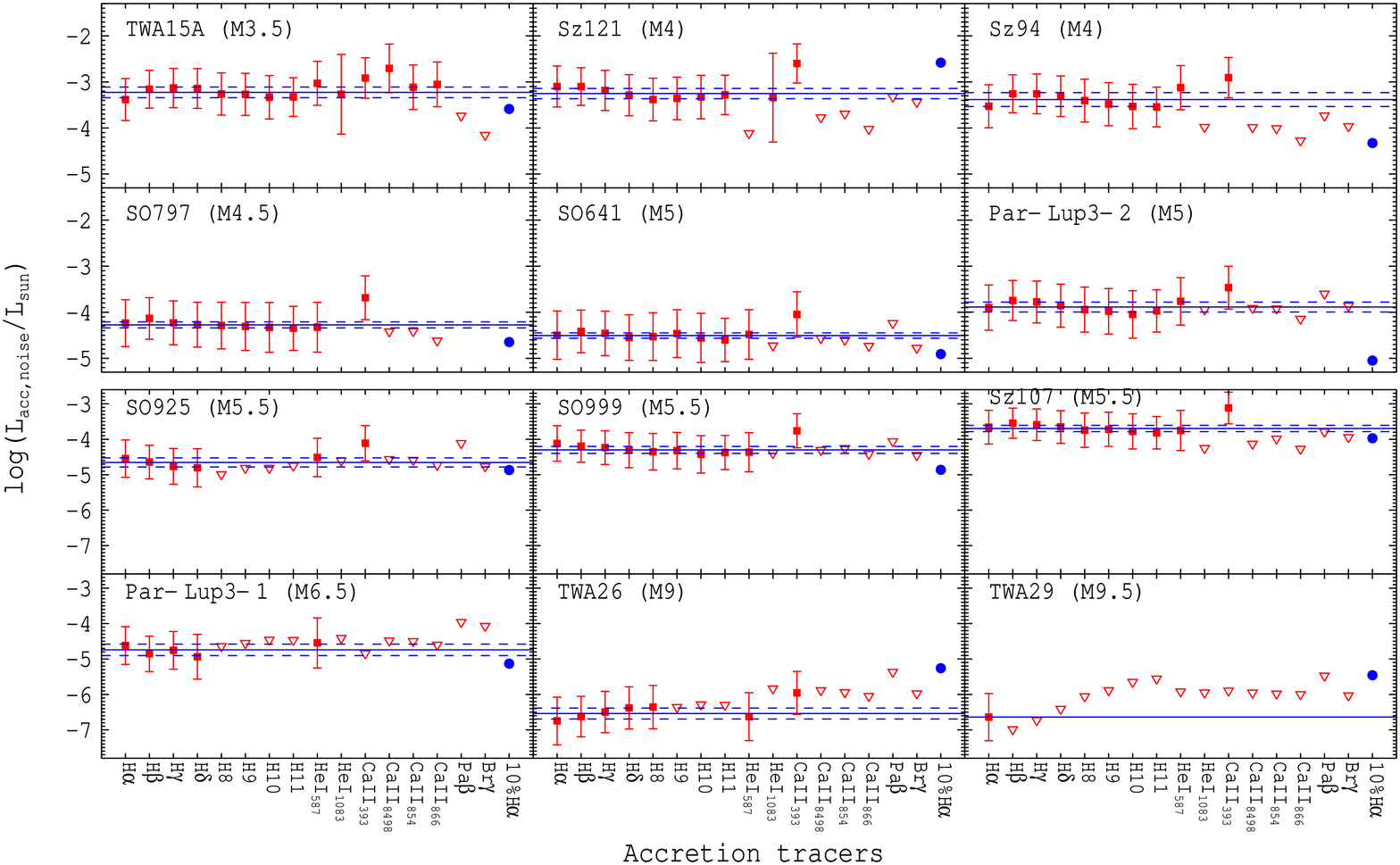} 
      \caption{Same as Fig.~\ref{Lacc_all_sources1}.}
         \label{Lacc_all_sources4}
   \end{figure*}

We include in Table~\ref{line_fluxes_table} the values of the observed Balmer jump, defined as the ratio between the flux at $\sim$360 nm and that at $\sim$400 nm. Typical values found in the literature for ClassIII YSOs range between $\sim$0.3 and 0.5 \citep{Herczeg08,Rigliaco12}. In ClassII YSOs, instead, the observed Balmer jump values are usually larger, up to $\sim$6 \citep{Hartigan91,Herczeg08, Rigliaco12}.  For the objects in our sample, this quantity ranges between $\sim$0.35 and $\sim$ 0.55 (see Table~\ref{line_fluxes_table}), with the exception of Sz122. The values of the Balmer jump ratio for the three BDs in the sample are not reported, because the SNR of the UVB spectrum of these sources is too low to estimate this quantity.


\section{Implications for mass accretion rates determination}
\label{sec::macc}
The physical parameter that is used to estimate the accretion activity in Class II YSOs is \macc, which is derived from $L_{\rm acc}$ and the stellar parameters using the following relation \citep{Hartmann98}:
\begin{equation}
\dot{M}_{\rm acc} = \left(1 - \frac{R_*}{R_{m}} \right)^{-1}\frac{L_{\rm acc}R_*}{GM_*} = \frac{L_{\rm acc}R_*}{0.8GM_*}\,,
\label{eq::macc}
\end{equation}
where the factor 0.8 is due to the assumption that the accretion flows arise from a magnetospheric radius $R_{\rm m}\sim5 R_*$ \citep{Shu}. 

When estimating $L_{\rm acc}$ in Class II YSOs with the direct method of UV-excess fitting \citep[e.g.][]{Valenti93}, the contribution to the continuum excess emission due to chromospheric activity is probably negligible and it is normally taken into account using as a template for the analysis a Class III YSO of the same SpT \citep[e.g.][]{Herczeg08,Rigliaco11a,Rigliaco12}. However, \lacc \ is often derived using spectral lines luminosity and appropriate empirical relations \citep[see e.g.][and references therein]{Muzerolle98,Natta04,Herczeg08}.

\begin{landscape}
\begin{table}			
\caption{\label{line_fluxes_table}Fluxes and equivalent widths of Balmer lines.}
\begin{tabular}{l|c|c|c|c|c|c|c|c|c|c}
\hline\hline
Name & H$\alpha$ & H$\beta$ & H$\gamma$ & H$\delta$ & H8 & H9 & H10 & H11 & 10\%H$\alpha$ & F$_{360}$/F$_{400}$ \\
\hline
TWA9A &    14.0 $\pm$     1.1 &     4.7 $\pm$     1.8 &     2.7 $\pm$     1.5 &     2.2 $\pm$     1.0 &     1.4 $\pm$     0.2 &     1.0 $\pm$     0.1 &    0.82 $\pm$    0.07 &    0.91 $\pm$    0.47 & 135.0 &  0.38 \\
\hbox{} & \hspace{0.4cm}$-$1.48 & \hspace{0.4cm}$-$0.68 & \hspace{0.4cm}$-$0.62 & \hspace{0.4cm}$-$0.59 & \hspace{0.4cm}$-$0.80 & \hspace{0.4cm}$-$0.83 & \hspace{0.4cm}$-$0.43 & \hspace{0.4cm}$-$0.65 & \hbox{} & \hbox{} \\
SO879 &     1.8 $\pm$     0.2 &    0.52 $\pm$    0.15 &    0.23 $\pm$    0.08 &    0.20 $\pm$    0.06 &    0.15 $\pm$    0.01 &    0.11 $\pm$    0.01 &    0.10 $\pm$    0.00 &   0.078 $\pm$   0.022 & 242.0 &  0.38 \\
\hbox{} & \hspace{0.4cm}$-$2.08 & \hspace{0.4cm}$-$1.17 & \hspace{0.4cm}$-$0.70 & \hspace{0.4cm}$-$1.12 & \hspace{0.4cm}$-$1.62 & \hspace{0.4cm}$-$1.57 & \hspace{0.4cm}$-$1.74 & \hspace{0.4cm}$-$0.77 & \hbox{} & \hbox{} \\
TWA6 &    36.4 $\pm$     4.0 &     8.4 $\pm$     2.1 &     5.1 $\pm$     3.9 &     3.3 $\pm$     0.7 &     2.0 $\pm$     0.1 &     1.5 $\pm$     0.1 &     1.4 $\pm$     0.1 &     1.3 $\pm$     0.7 & 362.0 &  0.41 \\
\hbox{} & \hspace{0.4cm}$-$4.17 & \hspace{0.4cm}$-$1.68 & \hspace{0.4cm}$-$1.02 & \hspace{0.4cm}$-$1.43 & \hspace{0.4cm}$-$1.47 & \hspace{0.4cm}$-$1.54 & \hspace{0.4cm}$-$1.67 & \hspace{0.4cm}$-$1.36 & \hbox{} & \hbox{} \\
TWA25 &    47.3 $\pm$     3.7 &    18.8 $\pm$     3.4 &    10.4 $\pm$     2.6 &     7.5 $\pm$     1.4 &     4.8 $\pm$     0.2 &     3.4 $\pm$     0.2 &     2.7 $\pm$     0.2 &     2.2 $\pm$     0.3 & 165.4 &  0.42 \\
\hbox{} & \hspace{0.4cm}$-$2.69 & \hspace{0.4cm}$-$1.94 & \hspace{0.4cm}$-$1.86 & \hspace{0.4cm}$-$1.81 & \hspace{0.4cm}$-$2.26 & \hspace{0.4cm}$-$1.97 & \hspace{0.4cm}$-$1.14 & \hspace{0.4cm}$-$1.15 & \hbox{} & \hbox{} \\
TWA14 &    15.4 $\pm$     1.1 &     3.6 $\pm$     0.5 &     2.1 $\pm$     0.5 &     1.4 $\pm$     0.2 &    0.87 $\pm$    0.02 &    0.63 $\pm$    0.02 &    0.50 $\pm$    0.02 &    0.41 $\pm$    0.07 & 262.2 &  0.45 \\
\hbox{} & \hspace{0.4cm}$-$5.77 & \hspace{0.4cm}$-$3.18 & \hspace{0.4cm}$-$3.03 & \hspace{0.4cm}$-$2.91 & \hspace{0.4cm}$-$3.30 & \hspace{0.4cm}$-$2.73 & \hspace{0.4cm}$-$1.28 & \hspace{0.4cm}$-$1.87 & \hbox{} & \hbox{} \\
TWA13B &    23.8 $\pm$     1.9 &     8.8 $\pm$     1.4 &     4.8 $\pm$     1.1 &     3.3 $\pm$     0.5 &     1.9 $\pm$     0.1 &     1.3 $\pm$     0.1 &    0.95 $\pm$    0.08 &    0.81 $\pm$    0.12 & 121.9 &  0.39 \\
\hbox{} & \hspace{0.4cm}$-$2.37 & \hspace{0.4cm}$-$1.94 & \hspace{0.4cm}$-$1.88 & \hspace{0.4cm}$-$1.75 & \hspace{0.4cm}$-$2.07 & \hspace{0.4cm}$-$1.65 & \hspace{0.4cm}$-$0.51 & \hspace{0.4cm}$-$0.94 & \hbox{} & \hbox{} \\
TWA13A &    85.7 $\pm$     5.5 &    30.2 $\pm$     4.0 &    17.0 $\pm$     3.4 &    10.6 $\pm$     1.3 &     4.8 $\pm$     0.1 &     3.0 $\pm$     0.1 &     2.3 $\pm$     0.1 &     1.7 $\pm$     0.2 & 287.8 &  0.45 \\
\hbox{} & \hspace{0.4cm}$-$7.58 & \hspace{0.4cm}$-$6.07 & \hspace{0.4cm}$-$5.48 & \hspace{0.4cm}$-$5.11 & \hspace{0.4cm}$-$4.30 & \hspace{0.4cm}$-$3.41 & \hspace{0.4cm}$-$1.28 & \hspace{0.4cm}$-$1.68 & \hbox{} & \hbox{} \\
TWA2A &    42.0 $\pm$     5.0 &    13.9 $\pm$     2.1 &     6.6 $\pm$     1.5 &     4.4 $\pm$     0.8 &     2.8 $\pm$     0.1 &     2.0 $\pm$     0.1 &     1.4 $\pm$     0.1 &     1.3 $\pm$     0.2 & 154.7 &  0.38 \\
\hbox{} & \hspace{0.4cm}$-$2.25 & \hspace{0.4cm}$-$1.88 & \hspace{0.4cm}$-$1.56 & \hspace{0.4cm}$-$1.36 & \hspace{0.4cm}$-$1.64 & \hspace{0.4cm}$-$1.38 & \hspace{0.4cm}$-$0.43 & \hspace{0.4cm}$-$0.84 & \hbox{} & \hbox{} \\
Sz122 &     6.4 $\pm$     0.8 &     3.2 $\pm$     0.3 &     2.3 $\pm$     0.6 &     1.5 $\pm$     0.4 &     1.2 $\pm$     0.1 &     1.0 $\pm$     0.1 &    0.86 $\pm$    0.05 &    0.59 $\pm$    0.08 & 660.3 &  0.74 \\
\hbox{} & \hspace{0.4cm}$-$6.76 & \hspace{0.4cm}$-$8.67 & \hspace{0.4cm}$-$10.00 & \hspace{0.4cm}$-$8.02 & \hspace{0.4cm}$-$11.57 & \hspace{0.4cm}$-$11.69 & \hspace{0.4cm}$-$10.02 & \hspace{0.4cm}$-$6.27 & \hbox{} & \hbox{} \\
TWA9B &     5.9 $\pm$     0.6 &     2.0 $\pm$     0.1 &     1.0 $\pm$     0.1 &    0.71 $\pm$    0.06 &    0.37 $\pm$    0.01 &    0.24 $\pm$    0.01 &    0.18 $\pm$    0.01 &    0.16 $\pm$    0.01 & 137.6 &  0.39 \\
\hbox{} & \hspace{0.4cm}$-$4.81 & \hspace{0.4cm}$-$4.83 & \hspace{0.4cm}$-$4.20 & \hspace{0.4cm}$-$3.50 & \hspace{0.4cm}$-$3.09 & \hspace{0.4cm}$-$2.49 & \hspace{0.4cm}$-$0.71 & \hspace{0.4cm}$-$1.54 & \hbox{} & \hbox{} \\
TWA15B &     6.8 $\pm$     0.3 &     2.0 $\pm$     0.1 &     1.1 $\pm$     0.1 &    0.72 $\pm$    0.06 &    0.38 $\pm$    0.01 &    0.27 $\pm$    0.02 &    0.19 $\pm$    0.01 &    0.16 $\pm$    0.01 & 157.7 &  0.45 \\
\hbox{} & \hspace{0.4cm}$-$8.96 & \hspace{0.4cm}$-$7.94 & \hspace{0.4cm}$-$7.19 & \hspace{0.4cm}$-$6.48 & \hspace{0.4cm}$-$6.20 & \hspace{0.4cm}$-$5.18 & \hspace{0.4cm}$-$1.60 & \hspace{0.4cm}$-$3.11 & \hbox{} & \hbox{} \\
TWA7 &    46.7 $\pm$     3.9 &    13.3 $\pm$     0.9 &     6.8 $\pm$     0.6 &     4.5 $\pm$     0.4 &     2.6 $\pm$     0.1 &     1.8 $\pm$     0.1 &     1.4 $\pm$     0.1 &     1.1 $\pm$     0.1 & 111.6 &  0.41 \\
\hbox{} & \hspace{0.4cm}$-$4.73 & \hspace{0.4cm}$-$4.53 & \hspace{0.4cm}$-$3.82 & \hspace{0.4cm}$-$3.36 & \hspace{0.4cm}$-$2.97 & \hspace{0.4cm}$-$2.36 & \hspace{0.4cm}$-$0.48 & \hspace{0.4cm}$-$1.68 & \hbox{} & \hbox{} \\
TWA15A &    10.4 $\pm$     0.6 &     3.0 $\pm$     0.2 &     1.8 $\pm$     0.1 &     1.3 $\pm$     0.1 &    0.62 $\pm$    0.02 &    0.47 $\pm$    0.02 &    0.33 $\pm$    0.02 &    0.25 $\pm$    0.03 & 250.3 &  0.55 \\
\hbox{} & \hspace{0.4cm}$-$12.43 & \hspace{0.4cm}$-$10.34 & \hspace{0.4cm}$-$10.80 & \hspace{0.4cm}$-$9.68 & \hspace{0.4cm}$-$6.33 & \hspace{0.4cm}$-$8.07 & \hspace{0.4cm}$-$3.36 & \hspace{0.4cm}$-$4.06 & \hbox{} & \hbox{} \\
Sz121 &     5.9 $\pm$     0.8 &     1.1 $\pm$     0.1 &    0.50 $\pm$    0.10 &    0.28 $\pm$    0.06 &    0.15 $\pm$    0.01 &    0.12 $\pm$    0.01 &    0.10 $\pm$    0.01 &   0.087 $\pm$   0.020 & 377.9 &  0.38 \\
\hbox{} & \hspace{0.4cm}$-$7.01 & \hspace{0.4cm}$-$5.06 & \hspace{0.4cm}$-$3.79 & \hspace{0.4cm}$-$2.90 & \hspace{0.4cm}$-$2.33 & \hspace{0.4cm}$-$2.39 & \hspace{0.4cm}$-$2.32 & \hspace{0.4cm}$-$2.02 & \hbox{} & \hbox{} \\
Sz94 &     2.3 $\pm$     0.2 &    0.75 $\pm$    0.05 &    0.42 $\pm$    0.04 &    0.27 $\pm$    0.02 &    0.14 $\pm$    0.01 &   0.087 $\pm$   0.006 &   0.064 $\pm$   0.006 &   0.047 $\pm$   0.008 & 188.9 &  0.39 \\
\hbox{} & \hspace{0.4cm}$-$5.67 & \hspace{0.4cm}$-$5.79 & \hspace{0.4cm}$-$5.49 & \hspace{0.4cm}$-$4.43 & \hspace{0.4cm}$-$3.91 & \hspace{0.4cm}$-$2.73 & \hspace{0.4cm}$-$2.66 & \hspace{0.4cm}$-$1.18 & \hbox{} & \hbox{} \\
SO797 &    0.16 $\pm$    0.02 &   0.035 $\pm$   0.003 &   0.016 $\pm$   0.002 & 0.0099 $\pm$ 0.0015 & 0.0057 $\pm$ 0.0006 & 0.0037 $\pm$ 0.0011 & 0.0030 $\pm$ 0.0011 & 0.0022 $\pm$ 0.0007 & 149.3 &  0.43 \\
\hbox{} & \hspace{0.4cm}$-$6.85 & \hspace{0.4cm}$-$6.83 & \hspace{0.4cm}$-$5.23 & \hspace{0.4cm}$-$3.89 & \hspace{0.4cm}$-$5.95 & \hspace{0.4cm}$-$3.71 & \hspace{0.4cm}$-$3.47 & \hspace{0.4cm}$-$1.48 & \hbox{} & \hbox{} \\
SO641 &   0.090 $\pm$   0.009 &   0.019 $\pm$   0.001 & 0.0097 $\pm$ 0.0010 & 0.0053 $\pm$ 0.0006 & 0.0033 $\pm$ 0.0003 & 0.0025 $\pm$ 0.0003 & 0.0018 $\pm$ 0.0003 & 0.0012 $\pm$ 0.0002 & 131.2 &  0.38 \\
\hbox{} & \hspace{0.4cm}$-$8.04 & \hspace{0.4cm}$-$9.08 & \hspace{0.4cm}$-$8.38 & \hspace{0.4cm}$-$5.42 & \hspace{0.4cm}$-$6.91 & \hspace{0.4cm}$-$5.98 & \hspace{0.4cm}$-$5.97 & \hspace{0.4cm}$-$1.74 & \hbox{} & \hbox{} \\
Par$-$Lup3$-$2 &     1.1 $\pm$     0.2 &    0.26 $\pm$    0.03 &    0.14 $\pm$    0.02 &   0.080 $\pm$   0.011 &   0.041 $\pm$   0.003 &   0.026 $\pm$   0.003 &   0.019 $\pm$   0.005 &   0.017 $\pm$   0.005 & 138.7 &  0.36 \\
\hbox{} & \hspace{0.4cm}$-$4.97 & \hspace{0.4cm}$-$4.88 & \hspace{0.4cm}$-$4.07 & \hspace{0.4cm}$-$2.67 & \hspace{0.4cm}$-$1.90 & \hspace{0.4cm}$-$1.38 & \hspace{0.4cm}$-$0.65 & \hspace{0.4cm}$-$1.39 & \hbox{} & \hbox{} \\
SO925 &   0.081 $\pm$   0.009 &   0.012 $\pm$   0.001 & 0.0050 $\pm$ 0.0011 & 0.0030 $\pm$ 0.0013 & $<$0.0012 & $<$0.0011 & $<$0.000975 & $<$0.000877 & 142.2 &  0.68 \\
\hbox{} & \hspace{0.4cm}$-$8.55 & \hspace{0.4cm}$-$6.81 & \hspace{0.4cm}$-$5.59 & \hspace{0.4cm}$-$4.29 & ... & ... & ... & ... & \hbox{} & \hbox{} \\
SO999 &    0.20 $\pm$    0.02 &   0.031 $\pm$   0.003 &   0.016 $\pm$   0.002 & 0.0091 $\pm$ 0.0020 & 0.0050 $\pm$ 0.0011 & 0.0036 $\pm$ 0.0006 & 0.0024 $\pm$ 0.0005 & 0.0021 $\pm$ 0.0007 & 147.1 &  0.45 \\
\hbox{} & \hspace{0.4cm}$-$8.70 & \hspace{0.4cm}$-$7.99 & \hspace{0.4cm}$-$7.80 & \hspace{0.4cm}$-$5.31 & \hspace{0.4cm}$-$4.23 & \hspace{0.4cm}$-$4.18 & \hspace{0.4cm}$-$2.43 & \hspace{0.4cm}$-$1.35 & \hbox{} & \hbox{} \\
Sz107 &     1.8 $\pm$     0.2 &    0.40 $\pm$    0.03 &    0.21 $\pm$    0.02 &    0.13 $\pm$    0.02 &   0.064 $\pm$   0.010 &   0.049 $\pm$   0.007 &   0.036 $\pm$   0.006 &   0.025 $\pm$   0.008 & 254.9 &  0.38 \\
\hbox{} & \hspace{0.4cm}$-$11.72 & \hspace{0.4cm}$-$11.77 & \hspace{0.4cm}$-$9.88 & \hspace{0.4cm}$-$6.91 & \hspace{0.4cm}$-$4.28 & \hspace{0.4cm}$-$4.23 & \hspace{0.4cm}$-$2.57 & \hspace{0.4cm}$-$2.07 & \hbox{} & \hbox{} \\
Par$-$Lup3$-$1 &    0.22 $\pm$    0.02 &   0.024 $\pm$   0.007 &   0.016 $\pm$   0.007 & 0.0073 $\pm$ 0.0060 & $<$0.0082 & $<$0.0066 & $<$0.0072 & $<$0.0055 & 140.0 &  1.43 \\
\hbox{} & \hspace{0.4cm}$-$15.78 & \hspace{0.4cm}$-$16.82 & \hspace{0.4cm}$-$29.99 & \hspace{0.4cm}$-$9.01 & ... & ... & ... & ... & \hbox{} & \hbox{} \\
TWA26 &   0.052 $\pm$   0.008 &   0.012 $\pm$   0.002 & 0.0085 $\pm$ 0.0016 & 0.0068 $\pm$ 0.0017 & 0.0038 $\pm$ 0.0007 & $<$0.0020 & $<$0.0022 & $<$0.0018 & 138.4 &  1.55 \\
\hbox{} & \hspace{0.4cm}$-$9.91 & \hspace{0.4cm}$-$27.07 & \hspace{0.4cm}$-$66.62 & \hspace{0.4cm}$-$35.39 & \hspace{0.4cm}$-$97.42 & ... & ... & ... & \hbox{} & \hbox{} \\
TWA29 &   0.019 $\pm$   0.003 & $<$0.0016 & $<$0.0014 & $<$0.0018 & $<$0.0021 & $<$0.0017 & $<$0.0028 & $<$0.0028 & 114.9 &  1.56 \\
\hbox{} & \hspace{0.4cm}$-$14.24 & ... & ... & ... & ... & ... & ... & ... & \hbox{} & \hbox{} \\
\hline
\end{tabular}
\tablefoot{Fluxes (10$^{-14}$ erg s$^{-1}$ cm$^{-2}$) are reported in the first rows for each objects with their errors, while equivalent widhts (\AA) in the second rows. The H$\alpha$ 10\% width is in km/s. The last column refers to the Balmer jump ratio, intened as the ratio between the flux at $\sim$360 nm to the flux at $\sim$400 nm. Upper limits are reported with $<$. }
\end{table}
\end{landscape}

\begin{landscape}
\begin{table}			
\caption{\label{line_fluxes_table_others}Fluxes and equivalent widths of Helium and Calcium lines.}
\begin{tabular}{l|c|c|c|c|c|c}
\hline\hline
Name & HeI$_{\lambda 587.6}$ & HeI$_{\lambda 1083}$ & CaII$_{\lambda 393.4}$ & CaII$_{\lambda 849.9}$ & CaII$_{\lambda 854.2}$ & CaII$_{\lambda 866.2}$ \\
\hline
TWA9A & 5.71e-15 $\pm$ 3.31e-15 & $<$1.10e-14 & 9.59e-14 $\pm$ 7.44e-16 & 4.53e-13 $\pm$ 3.21e-15 & 2.27e-13 $\pm$ 4.47e-15 & 2.59e-13 $\pm$ 4.36e-15 \\
\hbox{} & \hspace{0.4cm}$-$0.08 & ... & \hspace{0.4cm}$-$13.46 & \hspace{0.4cm}    0.16 & \hspace{0.4cm}    0.38 & \hspace{0.4cm}    0.46 \\
SO879 & 1.08e-15 $\pm$ 6.52e-16 & $<$5.63e-16 & 9.82e-15 $\pm$ 1.57e-16 & 4.04e-14 $\pm$ 3.13e-16 & 1.78e-14 $\pm$ 3.74e-16 & 2.26e-14 $\pm$ 3.65e-16 \\
\hbox{} & \hspace{0.4cm}$-$0.15 & ... & \hspace{0.4cm}$-$18.79 & \hspace{0.4cm}   -0.04 & \hspace{0.4cm}    0.28 & \hspace{0.4cm}    0.37 \\
TWA6 & 1.87e-14 $\pm$ 1.47e-14 & 2.99e-14 $\pm$ 9.97e-15 & 9.66e-14 $\pm$ 4.87e-15 & $<$3.54e-15 & $<$4.89e-15 & $<$2.85e-15 \\
\hbox{} & \hspace{0.4cm}$-$0.25 & \hspace{0.4cm}$-$0.39 & \hspace{0.4cm}$-$12.29 & ... & ... & ... \\
TWA25 & 2.39e-14 $\pm$ 1.14e-14 & 3.80e-14 $\pm$ 3.85e-14 & 2.46e-13 $\pm$ 1.60e-15 & 8.53e-13 $\pm$ 6.41e-15 & 3.86e-13 $\pm$ 9.47e-15 & 3.73e-13 $\pm$ 7.90e-15 \\
\hbox{} & \hspace{0.4cm}$-$0.19 & \hspace{0.4cm}$-$0.16 & \hspace{0.4cm}$-$20.04 & \hspace{0.4cm}    0.01 & \hspace{0.4cm}    0.23 & \hspace{0.4cm}    0.35 \\
TWA14 & 6.27e-15 $\pm$ 5.04e-15 & 1.61e-14 $\pm$ 5.51e-15 & 3.92e-14 $\pm$ 1.13e-15 & 1.54e-13 $\pm$ 1.29e-15 & 7.35e-14 $\pm$ 2.48e-15 & 6.42e-14 $\pm$ 1.05e-15 \\
\hbox{} & \hspace{0.4cm}$-$0.31 & \hspace{0.4cm}$-$0.44 & \hspace{0.4cm}$-$19.84 & \hspace{0.4cm}    0.02 & \hspace{0.4cm}    0.41 & \hspace{0.4cm}    0.49 \\
TWA13B & 1.66e-14 $\pm$ 7.83e-15 & 8.55e-14 $\pm$ 5.30e-14 & 1.22e-13 $\pm$ 7.24e-16 & 5.33e-13 $\pm$ 4.54e-15 & 2.13e-13 $\pm$ 6.06e-15 & 2.16e-13 $\pm$ 4.83e-15 \\
\hbox{} & \hspace{0.4cm}$-$0.24 & \hspace{0.4cm}$-$0.61 & \hspace{0.4cm}$-$21.91 & \hspace{0.4cm}    0.01 & \hspace{0.4cm}    0.22 & \hspace{0.4cm}    0.32 \\
TWA13A & 4.33e-14 $\pm$ 1.21e-14 & $<$1.73e-14 & 1.83e-13 $\pm$ 1.18e-14 & 6.50e-13 $\pm$ 5.87e-15 & 2.93e-13 $\pm$ 1.01e-14 & 3.18e-13 $\pm$ 7.42e-15 \\
\hbox{} & \hspace{0.4cm}$-$0.61 & ... & \hspace{0.4cm}$-$15.88 & \hspace{0.4cm}   -0.12 & \hspace{0.4cm}    0.15 & \hspace{0.4cm}    0.26 \\
TWA2A & 1.61e-14 $\pm$ 5.31e-15 & $<$3.93e-14 & 2.10e-13 $\pm$ 7.81e-16 & 1.23e-12 $\pm$ 1.05e-14 & 4.94e-13 $\pm$ 1.38e-14 & 3.84e-13 $\pm$ 9.01e-15 \\
\hbox{} & \hspace{0.4cm}$-$0.16 & ... & \hspace{0.4cm}$-$21.62 & \hspace{0.4cm}    0.08 & \hspace{0.4cm}    0.32 & \hspace{0.4cm}    0.42 \\
Sz122 & 2.93e-15 $\pm$ 6.09e-16 & 2.18e-15 $\pm$ 2.18e-15 & 1.88e-14 $\pm$ 7.91e-16 & $<$4.84e-16 & $<$6.36e-16 & $<$2.94e-16 \\
\hbox{} & \hspace{0.4cm}$-$0.44 & \hspace{0.4cm}$-$0.06 & \hspace{0.4cm}$-$19.58 & ... & ... & ... \\
TWA9B & 1.96e-15 $\pm$ 2.73e-16 & $<$3.74e-15 & 1.72e-14 $\pm$ 1.44e-16 & 1.15e-13 $\pm$ 1.33e-15 & 5.84e-14 $\pm$ 1.57e-15 & 3.64e-14 $\pm$ 9.35e-16 \\
\hbox{} & \hspace{0.4cm}$-$0.37 & ... & \hspace{0.4cm}$-$20.82 & \hspace{0.4cm}    0.02 & \hspace{0.4cm}    0.18 & \hspace{0.4cm}    0.32 \\
TWA15B & 2.63e-15 $\pm$ 5.10e-16 & 1.30e-14 $\pm$ 4.50e-15 & 1.51e-14 $\pm$ 3.17e-16 & 6.10e-14 $\pm$ 8.13e-16 & 2.69e-14 $\pm$ 1.36e-15 & 2.19e-14 $\pm$ 5.98e-16 \\
\hbox{} & \hspace{0.4cm}$-$0.69 & \hspace{0.4cm}$-$0.58 & \hspace{0.4cm}$-$27.41 & \hspace{0.4cm}   -0.20 & \hspace{0.4cm}    0.03 & \hspace{0.4cm}    0.19 \\
TWA7 & 1.66e-14 $\pm$ 2.82e-15 & $<$1.57e-14 & 1.42e-13 $\pm$ 8.57e-16 & 8.47e-13 $\pm$ 9.56e-15 & 3.90e-13 $\pm$ 1.19e-14 & 3.91e-13 $\pm$ 6.34e-15 \\
\hbox{} & \hspace{0.4cm}$-$0.35 & ... & \hspace{0.4cm}$-$25.00 & \hspace{0.4cm}   -0.10 & \hspace{0.4cm}    0.07 & \hspace{0.4cm}    0.22 \\
TWA15A & 4.40e-15 $\pm$ 6.80e-16 & 1.32e-14 $\pm$ 4.04e-15 & 1.79e-14 $\pm$ 8.08e-16 & 7.13e-14 $\pm$ 9.30e-16 & 3.11e-14 $\pm$ 1.59e-15 & 2.84e-14 $\pm$ 7.04e-16 \\
\hbox{} & \hspace{0.4cm}$-$1.05 & \hspace{0.4cm}$-$0.70 & \hspace{0.4cm}$-$19.29 & \hspace{0.4cm}   -0.31 & \hspace{0.4cm}   -0.02 & \hspace{0.4cm}    0.16 \\
Sz121 & $<$1.45e-16 & 3.50e-15 $\pm$ 3.22e-15 & 1.22e-14 $\pm$ 5.84e-16 & $<$1.30e-15 & $<$2.23e-15 & $<$7.06e-16 \\
\hbox{} & ... & \hspace{0.4cm}$-$0.11 & \hspace{0.4cm}$-$23.33 & ... & ... & ... \\
Sz94 & 1.12e-15 $\pm$ 1.41e-16 & $<$9.13e-16 & 5.64e-15 $\pm$ 1.53e-16 & $<$7.37e-16 & $<$9.93e-16 & $<$3.70e-16 \\
\hbox{} & \hspace{0.4cm}$-$0.59 & ... & \hspace{0.4cm}$-$18.92 & ... & ... & ... \\
SO797 & 2.92e-17 $\pm$ 7.82e-18 & $<$3.02e+02 & 2.48e-16 $\pm$ 2.09e-17 & $<$7.25e-17 & $<$1.11e-16 & $<$4.69e-17 \\
\hbox{} & \hspace{0.4cm}$-$0.35 & ... & \hspace{0.4cm}$-$20.54 & ... & ... & ... \\
SO641 & 2.12e-17 $\pm$ 3.81e-18 & $<$5.96e-17 & 1.00e-16 $\pm$ 2.47e-18 & $<$4.89e-17 & $<$6.87e-17 & $<$3.47e-17 \\
\hbox{} & \hspace{0.4cm}$-$0.65 & ... & \hspace{0.4cm}$-$16.82 & ... & ... & ... \\
Par$-$Lup3$-$2 & 3.00e-16 $\pm$ 7.21e-17 & $<$9.93e-16 & 1.39e-15 $\pm$ 4.54e-17 & $<$9.00e-16 & $<$1.25e-15 & $<$5.15e-16 \\
\hbox{} & \hspace{0.4cm}$-$0.46 & ... & \hspace{0.4cm}$-$9.30 & ... & ... & ... \\
SO925 & 1.98e-17 $\pm$ 3.89e-18 & $<$7.64e-17 & 8.55e-17 $\pm$ 7.46e-18 & $<$4.93e-17 & $<$7.12e-17 & $<$3.54e-17 \\
\hbox{} & \hspace{0.4cm}$-$0.79 & ... & \hspace{0.4cm}$-$18.79 & ... & ... & ... \\
SO999 & 2.70e-17 $\pm$ 8.90e-18 & $<$1.19e-16 & 2.05e-16 $\pm$ 3.06e-17 & $<$9.80e-17 & $<$1.65e-16 & $<$7.89e-17 \\
\hbox{} & \hspace{0.4cm}$-$0.42 & ... & \hspace{0.4cm}$-$15.56 & ... & ... & ... \\
Sz107 & 3.08e-16 $\pm$ 1.66e-16 & $<$5.19e-16 & 3.34e-15 $\pm$ 1.76e-16 & $<$5.03e-16 & $<$1.06e-15 & $<$3.70e-16 \\
\hbox{} & \hspace{0.4cm}$-$0.57 & ... & \hspace{0.4cm}$-$25.26 & ... & ... & ... \\
Par$-$Lup3$-$1 & 6.02e-17 $\pm$ 5.72e-17 & $<$3.71e-16 & $<$4.40e-17 & $<$1.97e-16 & $<$2.86e-16 & $<$1.58e-16 \\
\hbox{} & \hspace{0.4cm}$-$2.35 & ... & ... & ... & ... & ... \\
TWA26 & 1.89e-17 $\pm$ 9.96e-18 & $<$4.28e-16 & 6.20e-17 $\pm$ 1.82e-17 & $<$1.10e-16 & $<$1.69e-16 & $<$8.50e-17 \\
\hbox{} & \hspace{0.4cm}$-$1.88 & ... & \hspace{0.4cm}$-$56.20 & ... & ... & ... \\
TWA29 & $<$2.27e-17 & $<$9.49e-17 & $<$2.02e-17 & $<$2.59e-17 & $<$4.37e-17 & $<$2.73e-17 \\
\hbox{} & ... & ... & ... & ... & ... & ... \\
\hline
\end{tabular}

\tablefoot{Fluxes are reported in [erg s$^{-1}$ cm$^{-2}$] in the first rows, Equivalent Widths (EW) in [\AA] are reported in the second rows. Upper limits are shown with $<$. }
\end{table}
\end{landscape}

Several of the lines normally used for these studies are influenced by chromospheric activity (Sect.~\ref{sec::lines}), which should be estimated and subtracted from the line emission before computing \lacc. This procedure is not trivial, as it is difficult to properly disentangle the two emission processes in each object and each line. All the $L_{\rm acc} - L_{\rm lines}$ relations are always based on the non-corrected values of $L_{\rm line}$. This correction,  as shown in \citet{Ingleby11} and \citet{Rigliaco12}, can be important in objects with low $L_{\rm acc}$ and in particular in VLM stars. Therefore, the chromospheric emission acts as a systematic ``noise" which affects the $L_{\rm acc}$ measurements. 

In the following, we will characterize this effect using the $L_{\rm line}$ derived in Sect.~\ref{sec::lines} and the most recent $L_{\rm acc} - L_{\rm line}$ relations for Class II YSOs derived by \citet{Alcala13}. These relations have been derived using the same method as described in \citet{Rigliaco12}, where \lacc is obtained from the continuum excess emission and then compared with the \lline of the several emission line diagnostics in every object. In \citet{Alcala13} the sample is composed of 36 YSOs located in the Lupus-I and Lupus-III clouds, together with 8 additional YSOs located in the $\sigma$-Ori region from \citet{Rigliaco12}. These relations are not quantitatively different from those in the literature, but have significantly smaller uncertainties and have been derived for stars with similar properties than the Class III analyzed here. For each Class III object, we compute \lacc \ from a number of different lines, as described in the following. This provides a measurement of the ``noise'' introduced in the determination of \lacc \ from line luminosities in Class II. It represents a typical threshold for the determination of $L_{\rm acc}$ in Class II objects by chromospheric activity, assuming that this is approximately the same in the two different classes of objects, with and without ongoing accretion; we define it in the following as $L_{\rm acc,noise}$.

\subsection{Accretion luminosity noise}
\label{sec::laccnoise}
\citet{Alcala13} are using a sample of Class II YSOs in the Lupus star forming region observed with X-Shooter to refine the $L_{\rm acc} - L_{\rm line}$ relations. We adopt their relations to estimate the $L_{\rm acc,noise}$ for our Class III YSOs, using in particular the H$\alpha$, H$\beta$, H$\gamma$, H$\delta$, H8, H9, H10, H11, HeI ($\lambda\lambda$587.6, and 1083 nm), CaII ($\lambda$393 nm), CaII ($\lambda$849.8 nm), CaII ($\lambda$854.2 nm) and CaII ($\lambda$866.2 nm) lines. Moreover, we use the relation from \citet{Natta04} between \macc \  and the 10\% H$\alpha$ width to estimate \laccnoise \ from this indicator. 
 
We show in Fig.~\ref{Lacc_all_sources1}-\ref{Lacc_all_sources4} the values of $L_{\rm acc,noise}$ for every object obtained using the different indicators. The uncertainties on these values are dominated by the errors in the relations between \lline \ and \lacc. Upper limits for undetected lines are shown with triangles and indicate the 3$\sigma$ upper limits. Each Balmer line leads to values of $L_{\rm acc,noise}$ that are always in agreement with the other Balmer lines by less than $\sim$0.2 dex, and similarly the HeI$_{\lambda587.6}$ line almost in all cases. Solid lines represent the arithmetic mean of the Balmer and HeI$_{\lambda587.6}$ \laccnoise \ values, which is also reported in Table~\ref{sample_parameters}, while the dashed lines show the one sigma standard deviation. The HeI$_{\lambda1083}$ and the CaII lines, instead, are in various cases not in agreement with the result obtained using the Balmer and HeI$_{\lambda587.6}$ lines, with differences even larger than 0.6 dex. It should be considered that the exact value of the CaII IRT lines luminosity is subject to many uncertainties, due to the complicated procedure to estimate the excess luminosity (see Sect.~\ref{subsec::line_lum}). 

The Paschen and Brackett HI emission lines are not detected in our spectra, as pointed out in Sect.~\ref{subsec::em_lines}. We report in Fig.~\ref{Lacc_all_sources1}-\ref{Lacc_all_sources4} the 3$\sigma$ upper limits for \laccnoise \ obtained using the Pa$\beta$ and Br$\gamma$ line luminosities and the relations from \citet{Alcala13}. These values are always below the mean \laccnoise \ value obtained with the Balmer and HeI$_{\lambda587.6}$ lines. This implies that those lines are less sensitive to chromospheric activity than the Balmer lines.  

The 10\% H$\alpha$ width is the accretion indicator that leads to values of \laccnoise \ more discrepant from the mean (blue circles in Fig.~\ref{Lacc_all_sources1}-\ref{Lacc_all_sources4}). This clearly does not follow in more than 50\% of the cases the results obtained using the other indicators. This is not surprising as the H$\alpha$ width is mainly a kinematics measurement, contrary to \lline \  measurements. It is to be expected that the application of a method calibrated for accretion processes to chromospheric activity would result in inconsistencies. We know that the broadening of the H$\alpha$ line and the other accretion-related lines is due to the high-velocity infall of material in the accretion flows, while the intensity of the emission lines is due to emission from high temperature region. The latter can be either accretion shocks on the stellar surface or chromospheric emission. The fact that in our sample of non-accreting objects relations converting \lline \ to \lacc \ lead to similar results when using line fluxes, while the result is quite different when using line broadening seems to confirm that the H$\alpha$ 10\% width we detect is only due to thermal broadening in the chromosphere of these stars and not to gas flow kinematics associated with the accretion onto the central object.

   \begin{figure}
   \centering
   \includegraphics[width=0.5\textwidth]{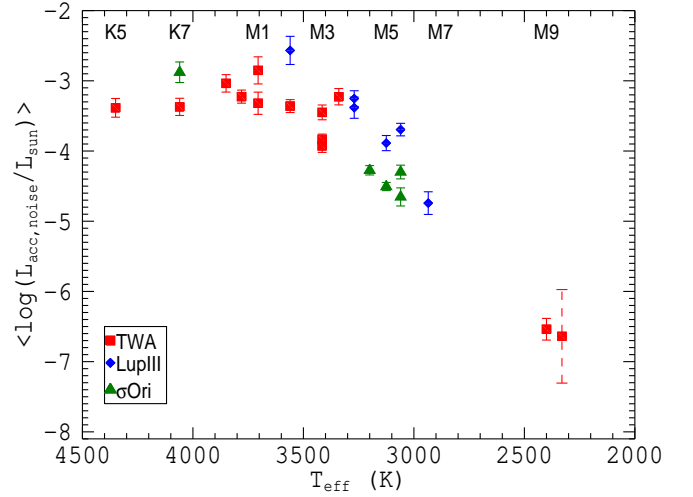} 
      \caption{Mean values of $\log (L_{\rm acc,noise}/L_\odot)$ obtained with different accretion diagnostics as a function of T$_{\rm eff}$. Error bars represent the standard deviation around the mean $\log (L_{\rm acc,noise}/L_\odot)$. These data should be intended as the noise in the values of $L_{\rm acc}$ due to chromospheric emission flux.}
         \label{Lacc_all_mean}
   \end{figure}

   \begin{figure}
   \centering
   \includegraphics[width=0.5\textwidth]{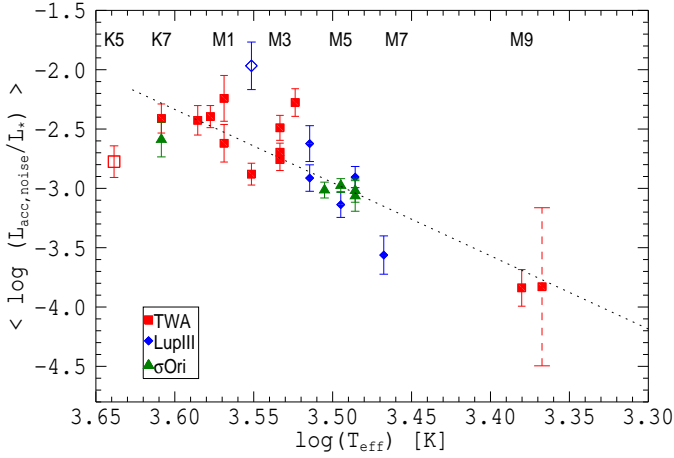} 
      \caption{Mean values of $\log (L_{\rm acc,noise}/L_*)$ obtained with different accretion diagnostics as a function of $\log$T$_{\rm eff}$. The dashed line is the best fit to the data, whose analytical form is reported in Eq.~(\ref{eq::fit}). Two objects (Sz122 and TWA9A) are excluded from the fit (empty symbols), as explained in the text.}
         \label{Lacc_Lstar_mean}
   \end{figure}

For all objects, the mean $L_{\rm acc,noise}$ value is always below $\sim 10^{-3} L_\odot$ and this value decreases monotonically with the SpT. In Fig.~\ref{Lacc_all_mean} these mean values of $\log (L_{\rm acc,noise}/L_\odot)$ obtained with the Balmer and HeI$_{\lambda587.6}$ lines are plotted as a function of the T$_{\rm eff}$ of the objects. The error bars on the plot represent the standard deviation of the derived values of \laccnoise\footnote{For the object TWA29, where only the H$\alpha$ line is detected, we report with a dashed line the error on the estimate of \laccnoise$/L_\odot$ from this line in Fig.~\ref{Lacc_all_mean}-\ref{Lacc_Lstar_mean}.}. These values should be intended as the noise in the \lacc \ values arising from the chromospheric activity. In Fig.~\ref{Lacc_Lstar_mean} we show the mean values of the logarithmic ratio \laccnoise/$L_*$ obtained using the Balmer and HeI$_{\lambda587.6}$ lines as a function of the T$_{\rm eff}$. Contrary to Fig.~\ref{Lacc_all_mean}, the quantity \laccnoise/L$_*$ is unbiased by uncertainties on distance values or by different stellar ages, leading to smaller spreads. We see that from the K7 objects down to the BDs, the values of log(\laccnoise/$L_*$) decrease with the T$_{\rm eff}$ of the YSOs. Fitting with a power-law the \laccnoise - T$_{\rm eff}$ relation, using only the objects in the range K7-M9.5 and excluding Sz122 (see Appendix~\ref{single_sources} for detail), we obtain the following analytical relation (dashed line in Fig.~\ref{Lacc_Lstar_mean}):

\begin{equation}
\log (L_{\rm acc,noise}/L_*) = (6.17\pm 0.53)\cdot \log T_{\rm eff} - (24.54\pm 1.88)
\label{eq::fit}
\end{equation}

The only clear deviation from the general trend, apart from Sz122, is the K5 YSO  TWA9A, which shows a value of $\log (L_{\rm acc,noise}/L_*)$ lower by $\sim$0.6 dex with respect to what should be expected by the extrapolation of the previous relation. Unfortunately, our sample is too small and we do not have other objects with earlier SpT to verify whether this low value is actually a different trend due to different chromospheric activity for earlier SpT YSOs or if the source is peculiar. There are also signatures of different chromospheric activity intensity among objects with the same SpT and located in the same region; for example, the two TW Hya M1 YSOs, which are two components of a binary system, thus coeval objects, have a spread in $\log (L_{\rm acc,noise}/L_*)$ of $\sim$0.5 dex.

Values of \lline \ and \lacc \ in Class II YSOs that are of the order of those estimated in this work should be considered very carefully, as the chromospheric activity could be an important factor of the excess luminosity in the line and could produce misleading results. 

  \begin{figure}
   \centering
   \includegraphics[width=0.5\textwidth]{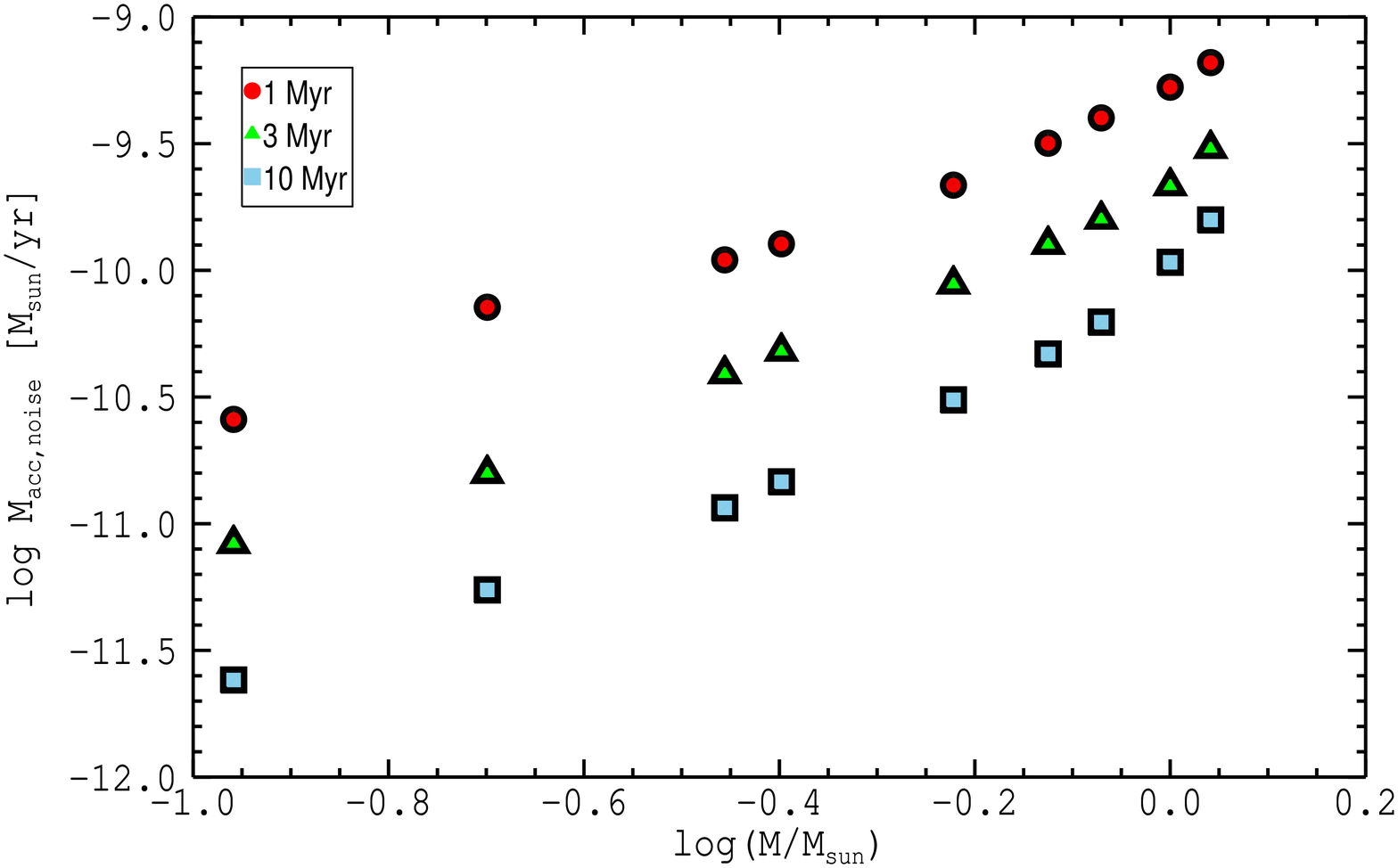} 
      \caption{$\log$\maccnoise \ as a function of $\log M_*$, with values of \maccnoise \ obtained using three different isochrones from \citet{Baraffe98} and the values of \laccnoise/$L_*$ derived from the fit in Eq.~(\ref{eq::fit}) at any $T_{\rm eff}$. Results using the 1 Myr isochrone are reported with filled circles, those using the 3 Myr isochrone with filled triangles, and those using the 10 Myr isochrone with filled squares.}
         \label{fig::Macc_noise_vs_M}
   \end{figure}

\subsection{Mass accretion rate noise}
\label{sec::macc_noise}
In this section, we determine what the typical ``chromospheric noise" on \macc \ would be when derived from indirect methods if the chromospheric emission is not subtracted before computing $L_{\rm line}$. We will refer to this quantity as \maccnoise.

The procedure used is the following. We select three isochrones (1, 3, and 10 Myr) from \citet{Baraffe98} models and nine different YSOs masses (0.11, 0.20, 0.35, 0.40, 0.60, 0.75, 0.85, 1.00, and 1.10 $M_\odot$), which correspond always to $T_{\rm eff}$ in the range 2500-4000 K, where our results are applicable. Then, for each $T_{\rm eff}$ we derive \laccnoise/$L_*$ using the fit reported in Eq.~(\ref{eq::fit}). Finally, we use Eq.~(\ref{eq::macc}), adopting the proper $R_*$, $M_*$ and $L_*$ at any age from the \citet{Baraffe98} tracks, in order to determine the typical \maccnoise \ at different ages as a function of $M_*$. The results are shown in Fig.~\ref{fig::Macc_noise_vs_M}, where a strong correlation between the two parameters is evident, with increasing \maccnoise \ with $M_*$. At the same time, the variation of \maccnoise \ with age at any given $M_*$ is large, up to 0.5 dex for differences of 2 Myr at the H-burning limit, but decreases with increasing $M_*$. Similar results are obtained when using other evolutionary models.

We derive a limit on the detectable \macc \ of $\sim 6.6\cdot10^{-10}$ $M_\odot$/yr for solar-mass, young (1 Myr) objects, decreasing to $2.5\cdot 10^{-12}$ $M_\odot$/yr for low-mass, older (10 Myr) objects. We report these \maccnoise \ values for the three isochrones analyzed in Table~\ref{tab::lim_macc}. 

   \begin{figure*}
   \centering
   \includegraphics[width=0.8\textwidth]{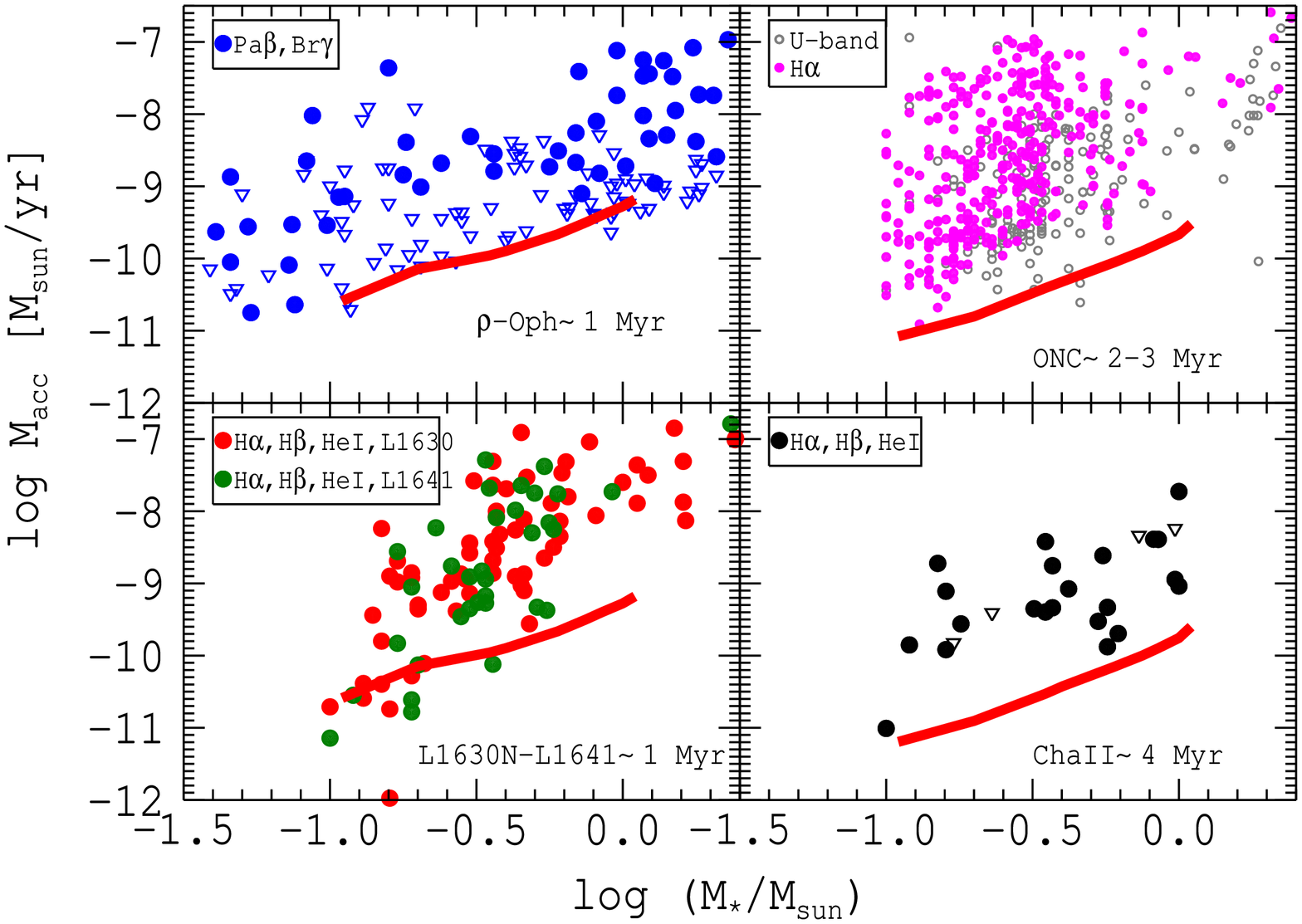} 
      \caption{$\log$\macc \ as a function of $\log M_*$ for Class II objects located in different star forming regions. Data for $\rho$Oph are from \citet{Natta06}, corrected for new distance measurements by \citet{Rigliaco11a}; ONC points are from \citet{Manara12}, data for L1630N and L1641 are from \citet{Fang09}, and those for ChaII from \citet{Biazzo12}. \macc \ values obtained with indirect methods are shown with coloured filled points, while measurements based on the direct $U$-band excess method are shown with grey empty points as a reference. Downwards triangles refer to upper limits. The thick red solid line is the lower limit to the measurements of \macc \ set by chromospheric activity in the line emission. We use the values for the correct isochrone according to the mean value of the age for each region, as reported on the plot.  }
         \label{fig::Macc_noise_check}
   \end{figure*}

\begin{table}
\caption{\label{tab::lim_macc}Values of $\log$\maccnoise \ at different $M_*$ and ages.}
\centering
\begin{tabular}{lccc}
\hline\hline
$M_*$ & \multicolumn{3}{c}{Age [Myr] } \\
\hbox{} [$M_\odot$] & 1 & 3 & 10 \\
\hline
    0.11  & $-$10.59   & $-$11.08   & $-$11.62 \\
    0.20  & $-$10.15   & $-$10.80   & $-$11.26 \\
    0.35   & $-$9.96   & $-$10.41   & $-$10.94 \\
    0.40   & $-$9.90   & $-$10.32   & $-$10.83 \\
    0.60   & $-$9.66   & $-$10.05   & $-$10.51 \\
    0.75   & $-$9.50    & $-$9.90   & $-$10.33 \\
    0.85   & $-$9.40    & $-$9.80   & $-$10.20 \\
    1.00   & $-$9.28    & $-$9.67    & $-$9.97 \\
    1.10   & $-$9.18    & $-$9.52    & $-$9.80 \\
\hline
\end{tabular}
\tablefoot{Values of log\maccnoise \ at different $M_*$ and different ages obtained using isochrones from \citet{Baraffe98} and the procedure explained in Sect.~\ref{sec::macc_noise}.}
\end{table}

\subsubsection{Comparison with literature data}
As discussed, \maccnoise \ is a lower limit to the values of \macc \ that can be derived from the luminosity of lines emitted by stellar chromospheres. In this section, we compare this chromospheric limit with estimates of \macc \ based on \lline \ from the literature. We show in Fig.~\ref{fig::Macc_noise_check} values of \macc \ as a function of $M_*$ obtained with optical or NIR emission lines as accretion diagnostic for objects located in four star forming regions with different ages: $\rho$-Ophiucus, the Orion Nebula Cluster, the L1630N and L1641 regions, and the Chameleon II region. We overplot with red solid lines the locii of \maccnoise \ obtained using the proper isochrone for the mean age of each region. In this way we do not address possible differences in \macc \ due to age spreads in these regions, but these spreads, if present, are $\lesssim$1-2 Myr \citep[see e.g.][]{Reggiani11}. The effect of a similar age spread on the \maccnoise \ threshold would be of $\sim$0.5 dex for low-mass objects and of $\sim$0.3 dex for solar-mass stars (see Fig.~\ref{fig::Macc_noise_vs_M}).

We consider in this analysis \macc \ values obtained using optical emission lines. In Fig.~\ref{fig::Macc_noise_check} we show the values for the $\sim$1 Myr old regions L1630N and L1641 \citep{Fang09}, where the H$\alpha$, H$\beta$, and HeI lines were used, those for the $\sim$ 2-3 Myr old Orion Nebula Cluster \citep{Manara12} obtained with photometric narrow-band H$\alpha$ luminosity estimates\footnote{We report also the values of \macc \ obtained through $U$-band excess by \citet{Manara12}, which are shown with empty symbols as a comparison.}, and those obtained with the H$\alpha$, H$\beta$, and HeI lines for the $\sim$4 Myr old Chameleon II region \citep{Biazzo12}. In all these cases, the vast majority of datapoints are found, as expected, well above the \maccnoise \ threshold, confirming that chromospheric contribution to the line emission is negligible for strongly accreting YSOs. Nevertheless, there are a few (11) objects in L1630N and L1641, where measured values of \macc \ are smaller than the expected \maccnoise. One possibility to explain this result is variable accretion in these objects, which has been found to vary as much as 0.4 dex over time-scales of 1 year \citep{Costigan12}. Still, this does not explain why only lower mass object happen to be below the threshold. In any case, these points should be considered with caution, as the measured \lline \ could be completely due to chromospheric emission, leading to erroneous estimates of \macc.

For the $\sim$1 Myr old $\rho$ Ophiucus region we consider the \lline \ derived through Pa$\beta$ and Br$\gamma$ lines \citep{Natta06}, corrected for a more recent estimate of the distance \citep[see][]{Rigliaco11a}. As we noted in Sect.~\ref{subsec::em_lines} and~\ref{sec::laccnoise}, Pa$\beta$ and Br$\gamma$ lines are not detected in our Class III YSOs spectra. In Fig.~\ref{fig::Macc_noise_check}, all the detections (blue filled circles) are located well above the \maccnoise \ locus, while the upper limits (blue open triangles) are distributed also at the edge of the \maccnoise \ threshold. This confirms the validity of these NIR lines as good tracers of accretion and the fact that they are most likely less subject to chromospheric noise than the Balmer and HeI$_{\lambda587.6}$ lines.


\section{Conclusion}
\label{sec::conclusion}
In this paper, we presented the analysis of 24 diskless, hence non-accreting Class III YSOs, observed with the broad-band, medium-resolution, high-sensitivity VLT/X-Shooter spectrograph. The targets are located in three nearby star forming regions (Lupus III, $\sigma$ Ori and TW Hya) and have SpT in the range from K5 to M9.5. We have checked the SpT classifications, using both spectral indices and broad molecular bands. Moreover, using the flux calibrated spectra, we have derived the stellar luminosity. Then, we have analyzed the emission lines related to accretion processes in accreting objects that are present in these spectra and, from their luminosities, we have studied the implications of chromospheric activity for \macc \ determination in accreting (Class II) objects. This has been done by deriving the parameter \laccnoise, which is the systematic ``noise" introduced by chromospheric emission in the measurements of \lacc \ from line emission in Class II objects. For this analysis we assumed a similar chromospheric activity in the two classes of objects. 

Our main conclusions are:
\begin{enumerate}
\item All hydrogen recombination emission lines of the Balmer series are detected in our sample of Class III YSOs spectra when the S/N is high enough. On the opposite, Paschen and Brackett series lines are not detected in emission, and they are significantly weaker, when compared to Balmer lines, then in Class II objects. The chromospheric ``noise'' in these lines is lower than in the optical lines (see Fig.~\ref{Lacc_all_sources1}-\ref{Lacc_all_sources4}), and they are very good tracers of accretion in low-mass, low accretion rates objects.
\item Using Balmer and HeI$_{\lambda587.6}$ lines and the calibrated relations between $L_{\rm line}$ and $L_{\rm acc}$ from the literature, we derive \laccnoise \ values always in good agreement among all the lines.
\item Calcium emission lines in the NIR spectral range ($\lambda\lambda$ 849.8, 854.2, 866.2 nm) are detected in 11 objects (45\% of the sample) superposed on the photospheric absorption lines. This results in a more complicated line flux measurement than for other lines. Their behavior with respect to the hydrogen line luminosities is different; in particular, the values of $L_{\rm acc,noise}$ obtained using these lines are often not in good agreement with those obtained using Balmer and HeI$_{\lambda587.6}$ lines. 
\item The mean values of $L_{\rm acc,noise}$ for the objects in our sample are smaller than $\sim$10$^{-3} L_\odot$  and have a clear dependence with T$_{\rm eff}$ for K7-M9.5 objects. Therefore, \lacc \ of this order or smaller measured in Class II objects using line luminosity as a proxy of accretion may be significantly overestimated if the chromospheric contribution to the line luminosity is not taken into account.
\item Our results show that the ``noise" due to chromospheric activity on the estimate of \macc \ in Class II YSOs obtained using secondary indicators for accretion has a strong dependence with $M_*$ and age. Typical values of log(\maccnoise) for M-type YSOs are in the range from $\sim-9.2$ for solar-mass young (1 Myr) objects to $-$11.6 $M_\odot$/yr for low-mass, older (10 Myr) objects. Therefore, derived accretion rates below this threshold should be treated with caution as the line emission may be dominated by chromospheric activity. 

\end{enumerate}


\begin{acknowledgements}
 C.F.M. acknowledges the PhD fellowship of the International Max-Planck-Research School. We thank Aleks Scholz, Greg Herczeg, and Luca Ricci for insightful discussions.\\
 This research has made use of the SIMBAD database and of the VizieR catalogue access tool, operated at CDS, Strasbourg, France.
\end{acknowledgements}


\appendix
\section{Comments on individual objects}
\label{single_sources}

\subsection{Sz94}
Sz94 was discovered as an H$\alpha$ emitting star in the survey by Schwarz (1977). Since then, it has been considered as a PMS star in reviews and investigations \citep{Krautter92, Hughes94, Lupus, Mortier11}, but nothing was mentioned about the presence of the lithium absorption line at 670.8 nm. Based on its spectral energy distribution, Sz94 has been classified as a Class III IR YSO by \citet{Merin08}. For all these reasons, we included the star in our programme as a Class III template of M4 SpT. Notwithstanding the high quality of the X-Shooter data in terms of S/N and resolution, the Li~I $\lambda$ 670.8 nm line is not present in the spectrum, and we determine an upper limit of EW$_{\rm Li}<0.1$\AA. The question then rise whether lithium may have already been depleted in the star. However, lithium is significantly depleted by large factors only after several tens of Myr, inconsistently with the average age of a few Myr of the Lupus members. Analysis on the radial velocity of this object show that its value is in the range of typical values for Lupus sources \citep{Stelzer13b}. We consider here the source as a PMS given its position in the HRD, the H$\alpha$ emission and the radial velocity measurement, but more analysis should be carried out to confirm its PMS status. 

\subsection{Sz122}
Sz122 is a Class III object classified with {\it Spitzer} \citep{Merin08}. All the lines of this object appear very broadened. Nevertheless, the spectrum cannot be fitted with synthetic spectra broadened at reasonable values of $v sini$, meaning that this is not a single fast rotator. We think, therefore, that this object is a binary system. This could explain the very faint LiI line (EW$_{\rm Li} <$ 0.25 \AA) and the lack of emission in the CaII IRT absorption features, that is usually found in other early M-type objects of our sample. Even the presence of a broad H$\alpha$ line (10\% width $>$ 600 km/s) can be explained by the presence of two objects. We do not consider this object in the analysis of the implications of chromospheric activity on \macc \ measurements, because the effect of its binarity cannot be accounted for.

\subsection{Sz121}
Similarly to Sz122, the lines of this object appear very broadened. Also in this case, the object is classified with {\it Spitzer} as a Class III \citep{Merin08}. We can fit the spectrum with a synthetic spectrum of the same T$_{\rm eff}$, log$g$=4.0 and $vsini$ = 70 km/s. This value of $vsini$ is rather high for a typical M4-type YSO. If true, it would be an extreme ultra-fast rotator.  The large 10\% H$\alpha$ width ($\sim$380 km/s) is then due due to its very high rotational velocity. Another possibility, is that this object may also be an unresolved spectroscopic binary.

\subsection{TWA6}
This object has been discovered and classified as a young non-accreting object by \citet{Webb99}. It is a member of the TWA association and it is a known fast rotator ($v sin i$=72 km/s, \citet{Skelly08}).  We measure a 10\% H$\alpha$ width of 362 km/s. This is larger than the threshold to distinguish accreting and non-accreting objects proposed by \citet{WB03}. Nevertheless, this object can be considered a Class III YSO for the small EW$_{H\alpha}$, the small Balmer jump and, in particular, the IR classification. For these reasons we consider this object in our analysis.


\section{NIR spectral indices}
\label{NIR_ind_app}
Spectral indices in the NIR part of the spectrum are particularly useful to classify VLM stars and BDs, which emit most of their radiation in this spectral region. Therefore, various analyses to calibrate reliable NIR indices to classify late M-type objects and BDs have been carried out in the past \citep[e.g.][]{Kirkpatrick99}. With our sample, we are able to verify the validity of various NIR indices for M-type YSOs, comparing the SpT obtained through optical spectroscopy (Sect.~\ref{sec::source_class}). We consider in this analysis different NIR indices which have been calibrated using either dwarfs or young stars (subgiants) for objects with SpT M or later. In particular, \citet{Rojas-Ayala12} calibrated the $H_2 O-K2$ index using a sample of M dwarfs, and derived a relation valid for the whole M class; \citet{Allers07} calibrated on a sample of young BDs and dwarfs the gravity independent spectral index $H_2O$, which is valid for objects with SpT in the range M5-L0; \citet{Testi01} and \citet{Testi09} proposed various spectral indices (sH$_2$O$^J$, sH$_2$O$^K$, sKJ, sHJ, sH$_2$O$^{H1}$, sH$_2$O$^{H2}$, I$_J$, I$_H$, and I$_K$) to classify M-, L-, and T-type BDs, calibrating those on a large sample of dwarfs; finally, \citet{Scholz12} used a sample of VLM YSOs to calibrate the HP index, which is valid for objects with SpT in the range M7-M9.5. We report those indices in Table~\ref{tab::NIR_indices}.

   \begin{figure*}
   \centering
   \includegraphics[width=\textwidth]{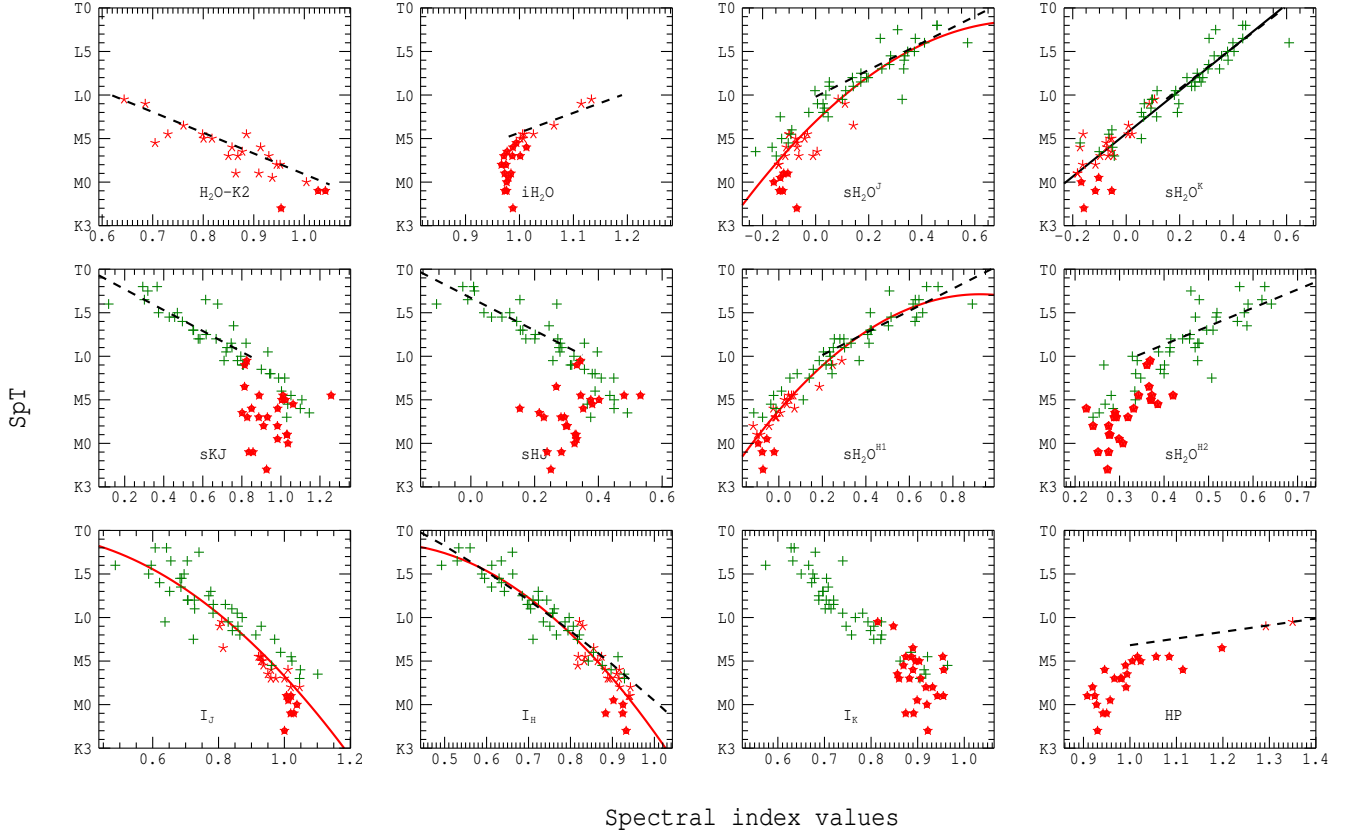} 
      \caption{Spectral type of the objects as a function of the spectral index values obtained with different NIR indices (Table~\ref{tab::NIR_indices}). Red symbols are values from this work, while green crosses are from \citet{Testi01} and \citet{Testi09}. Red stars are used for objects with SpT in the range of validity of the index, while red pentagons for those not in the range of validity of the index. Dashed lines are the best fit from the reference cited in Table~\ref{tab::NIR_indices} for each index, when available. Black and red solid lines are the best fits (black = linear, red = polynomial) from this work, considering both our sample and the values from the literature. }
         \label{fig::NIR_indices}
   \end{figure*}

Fig.~\ref{fig::NIR_indices} shows the results of this analysis: the SpTs obtained in Sect.~\ref{sec::source_class} are reported as a function of each spectral index for our objects (red crosses when SpTs are in the range of validity of an index and red circles when not). For the spectral indices derived in \citet{Testi01} and \citet{Testi09}, we report also the data for the sample of L- and M-type dwarfs they adopted in the analysis (green crosses).  We stress that the latter sample should be considered with cautions, as gravity dependent spectral indices calibrated using samples of dwarfs may not be reliable for YSOs, which are subgiants. The published relations between the SpT and the spectral indices in the range of validity of each index are shown as black dashed lines\footnote{No published relations are available for the indices I$_J$ and I$_K$}. 

We see that the indices sH$_2$O$^{H2}$, sKJ, sHJ, and I$_K$ cannot be used to classify YSOs of M-type class. On the opposite, we find a good correlation with the indices sH$_2$O$^K$, sH$_2$O$^J$, sH$_2$O$^{H1}$, I$_J$, and I$_H$. For these indices we decide to fit altogether our objects and those from the literature in all cases, using either a linear or a second degree polynomial fit. We show in Fig.~\ref{fig::NIR_indices} the best fit obtained, with a black solid line in case of linear fit and a red solid line when the best fit is polynomial. Using these relations, we derive the SpT for each of our YSOs, and we report these results in Table~\ref{tab::NIR_indices_results}. We thus propose new SpT-spectral index relations for the following five indices from \citet{Testi01} and \citet{Testi09}. The first three are valid for objects with SpT later and equal to M1:

\begin{equation}
{\rm SpT-code} = 0.55 + 2.48 \cdot sH_2O^K 
\end{equation}
\begin{equation}
 {\rm SpT-code} = 0.38 + 2.89 \cdot sH_2O^{H1} -1.56 \cdot (sH_2O^{H1} )^2
\end{equation}
\begin{equation}
{\rm SpT-code} = 1.11 + 3.92\cdot I_H -5.35 \cdot (I_H)^2
\end{equation}

The folloewing two indices are valid for objects with SpT later and equal to M2:
\begin{equation}
{\rm SpT-code} = 0.70 + 2.90 \cdot sH_2O^J -1.78 \cdot (sH_2O^J)^2
\end{equation}
\begin{equation}
{\rm SpT-code} = 1.83 + 1.08 \cdot I_J - 2.59 \cdot (I_J)^2
\end{equation}
where the SpT are coded in the following way: M0 $\equiv$ 0.0, M9 $\equiv$ 0.9, L5 $\equiv$ 1.5 and a variation of 0.1 corresponds to a step of one subclass. For these spectral indices we conclude that results for objects with SpT in the nominal range of validity of each index are reliable within a typical uncertainty of $\sim$ one subclass. 

A good correlation is also found for the spectral index $H_2 O-K2$; using the analytical relation between SpT and spectral index from the literature, we obtain for 13 out of 22 objects with M SpT results that are compatible within one sub-class with the correct SpT (see Table~\ref{tab::NIR_indices_results}). The differences can be due to the non perfect telluric removal in the first interval of interest of this index. Moreover, we confirm that the H$_2$O index is valid for YSOs with SpT in the range M5-M9.5, finding an agreement within one sub-class for all our objects (see Table~\ref{tab::NIR_indices_results}). Regarding the HP index, we confirm that it is not valid for YSOs with SpT earlier than M7, and we observe that the SpT obtained with this index confirm those from literature for our two later SpT objects (see Table~\ref{tab::NIR_indices_results}).

\begin{table*}[!]
\caption{\label{tab::NIR_indices}NIR spectral indices analysed in Appendix~\ref{NIR_ind_app}.}
\centering
\begin{tabular}{lcccl}
\hline\hline
Index&Range of validity&Numerator [nm]&Denominator [nm] &Reference\\
\hline
H$_2$O-K2          & M0-M9     & (2070-2090) / (2235-2255) & (2235-2255) / (2360-2380)  & \citet{Rojas-Ayala12} \\
I$_J$       & M0-T9     & (1090-1130) + (1330-1350) & 2$\cdot$(1265-1305)  & \citet{Testi09} \\
I$_H$       & M0-T9     & (1440-1480) + (1760-1800) & 2$\cdot$(1560-1600)  & \citet{Testi09} \\
I$_K$       & M0-T9     & (1960-1990) + (2350-2390) & 2$\cdot$(2120-2160)  & \citet{Testi09} \\
H$_2$O         & M5-L5     & 1550-1560 & 1492-1502  & \citet{Allers07} \\
sHJ         & L0-L9     & (1265-1305) - (1600-1700) & 0.5$\cdot$[(1265-1305) + (1600-1700)]  & \citet{Testi01} \\
sKJ        & L0-L9     & (1265-1305) - (2120-2160) & 0.5$\cdot$[(1265-1305) + (2120-2160)]  & \citet{Testi01} \\
sH$_2$O$^J$        & L0-L9     & (1265-1305) - (1090-1130) & 0.5$\cdot$[(1265-1305) + (1090-1130) ] & \citet{Testi01} \\
sH$_2$O$^{H1}$        & L0-L9     & (1600-1700) - (1450-1480) & 0.5$\cdot$[(1600-1700) + (1450-1480)]  & \citet{Testi01} \\
sH$_2$O$^{H2}$        & L0-L9     & (1600-1700) - (1770-1810) & 0.5$\cdot$[(1600-1700) + (1770-1810)]  & \citet{Testi01} \\
sH$_2$O$^K$        & L0-L9     & (2120-2160) - (1960-1990) & 0.5$\cdot$[(2120-2160) + (1960-1990)]  & \citet{Testi01} \\
HIP & M7-M9.5 & (1675-1685) & (1495-1505) & \citet{Scholz12} \\
\hline
\end{tabular}
\end{table*}


\clearpage

\Online
\section{On-line material}
\label{app::spectra}

\subsection{NIR spectra}

   \begin{figure*}[b!]
   \centering
   \includegraphics[width=\textwidth]{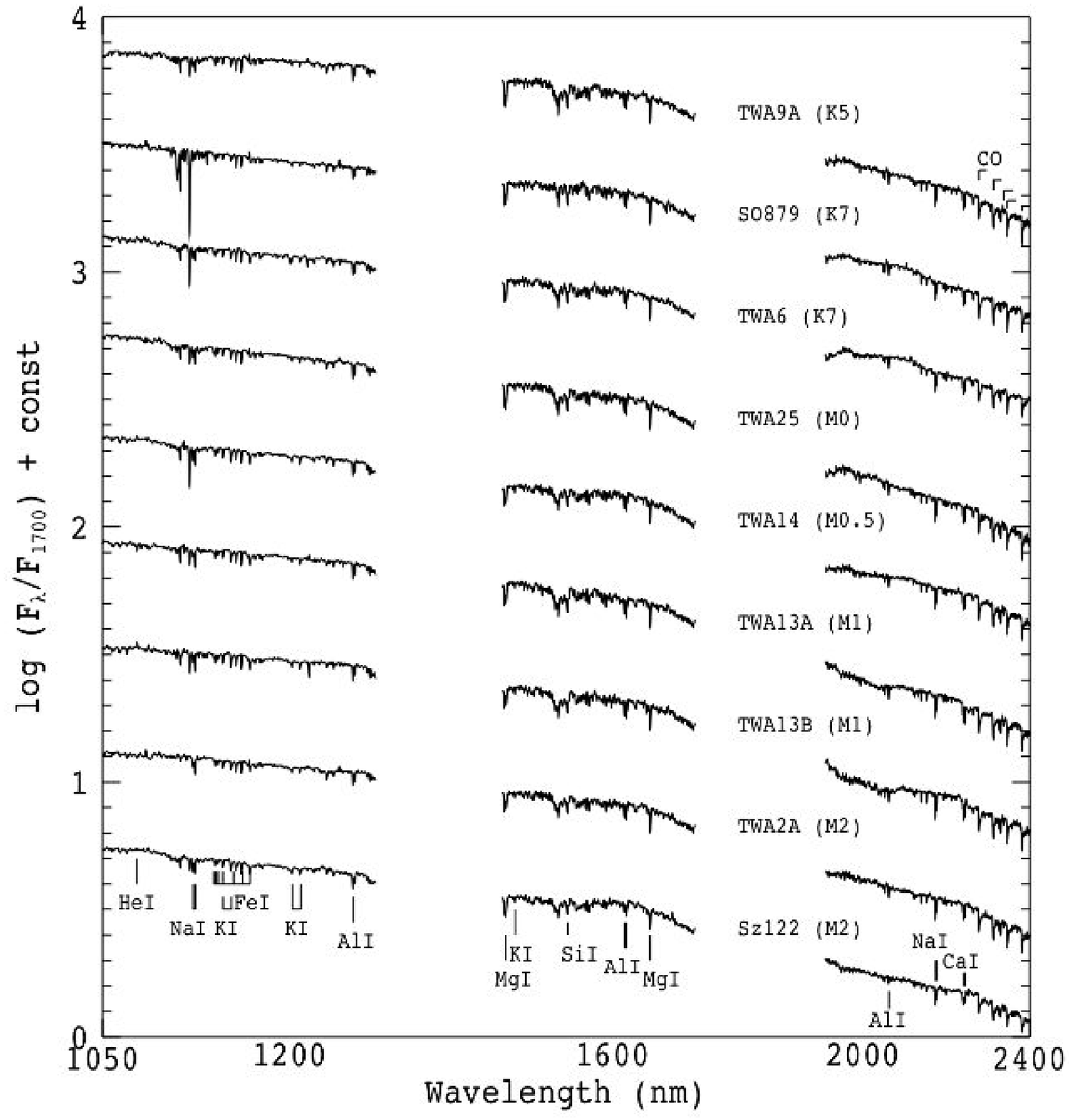}	 
      \caption{Spectra of Class III YSOs with spectral type earlier than M3 in the NIR arm. All the spectra are normalized at 1700 nm and offset in the vertical direction by 0.5 for clarity. The spectra are also smoothed to the resolution of 2000 at 2000 nm to make easier the identification of the features.}
         \label{spec_NIR}
   \end{figure*}

   \begin{figure*}[!]
   \centering
   \includegraphics[width=\textwidth]{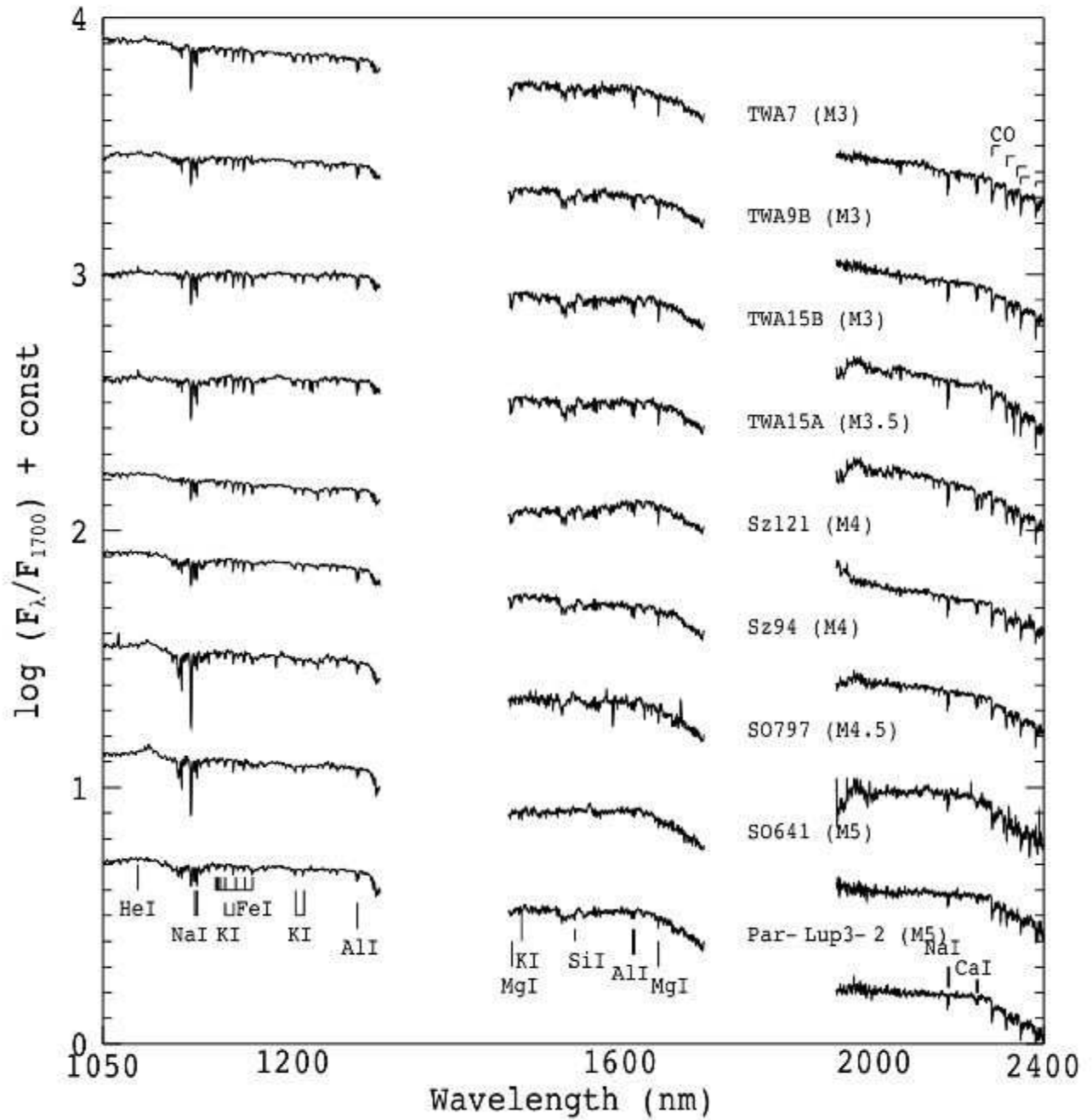}
      \caption{Same as Fig.~\ref{spec2_NIR}, but for spectral types between M3 and M5. }
	\label{spec2_NIR}
   \end{figure*}

   \begin{figure*}[!]
   \centering
   \includegraphics[width=\textwidth]{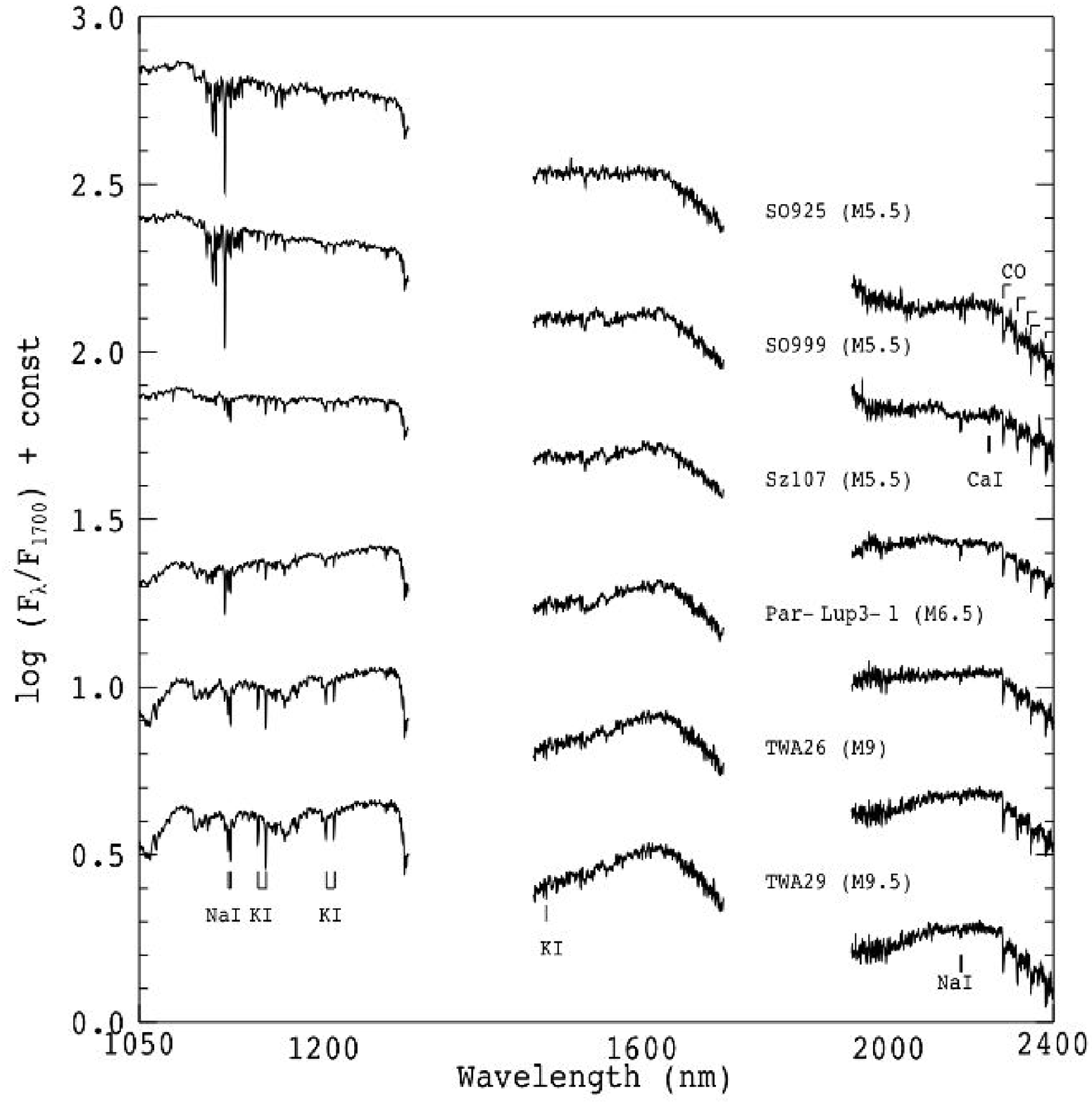}
      \caption{Same as Fig.~\ref{spec3_NIR}, but for spectral types later than M5. }
	\label{spec3_NIR}
   \end{figure*}

\clearpage

\subsection{UVB spectra}

   \begin{figure*}
   \centering
   \includegraphics[width=0.9\textwidth]{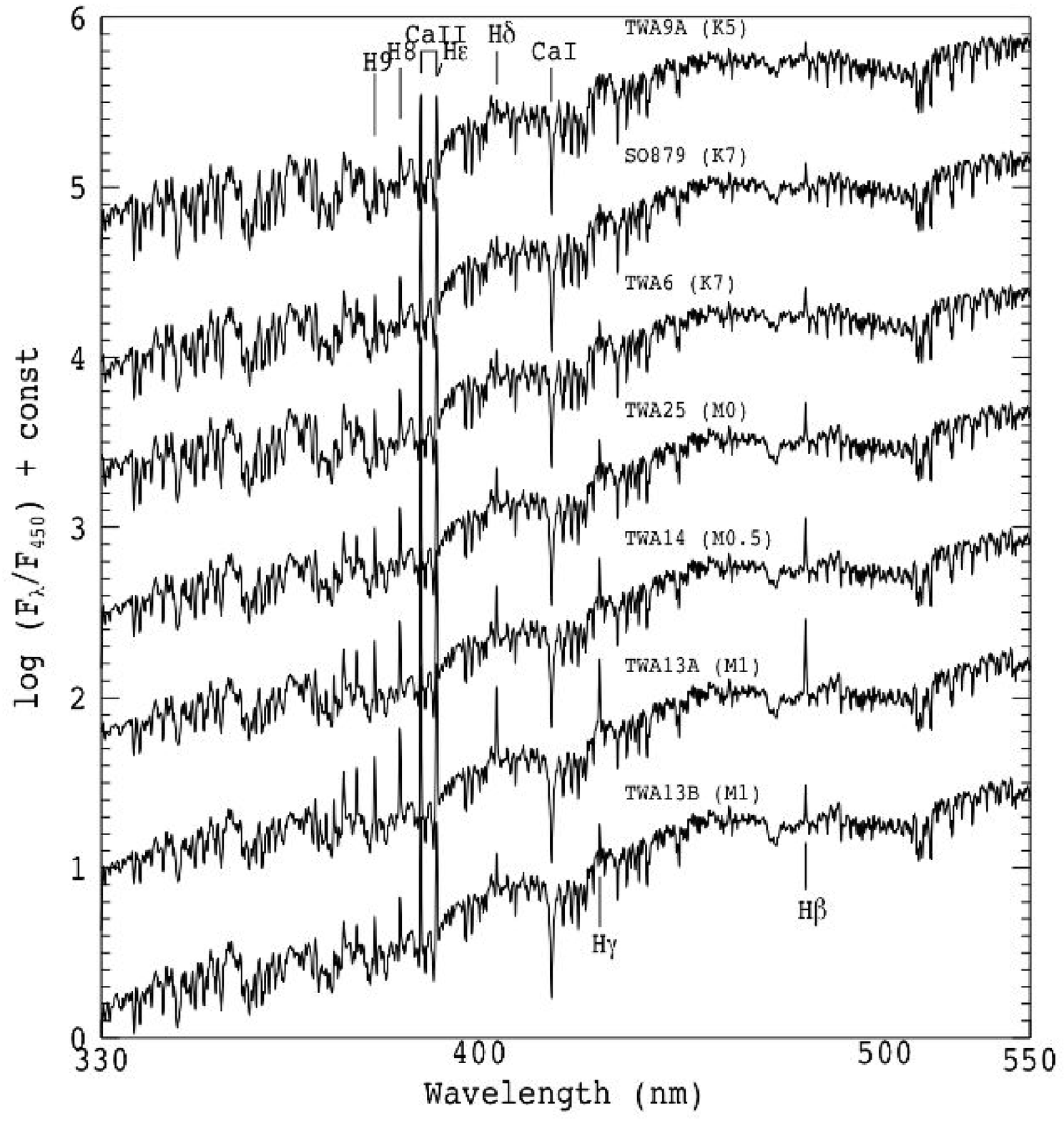}	 
      \caption{Spectra of Class III YSOs with spectral type earlier than M2 in the UVB arm. All the spectra are normalized at 450 nm and offset in the vertical direction for clarity. The spectra are also smoothed to the resolution of 1500 at 400 nm to make easier the identification of the features.}
         \label{spec_UVB}
   \end{figure*}

   \begin{figure*}[!]
   \centering
   \includegraphics[width=\textwidth]{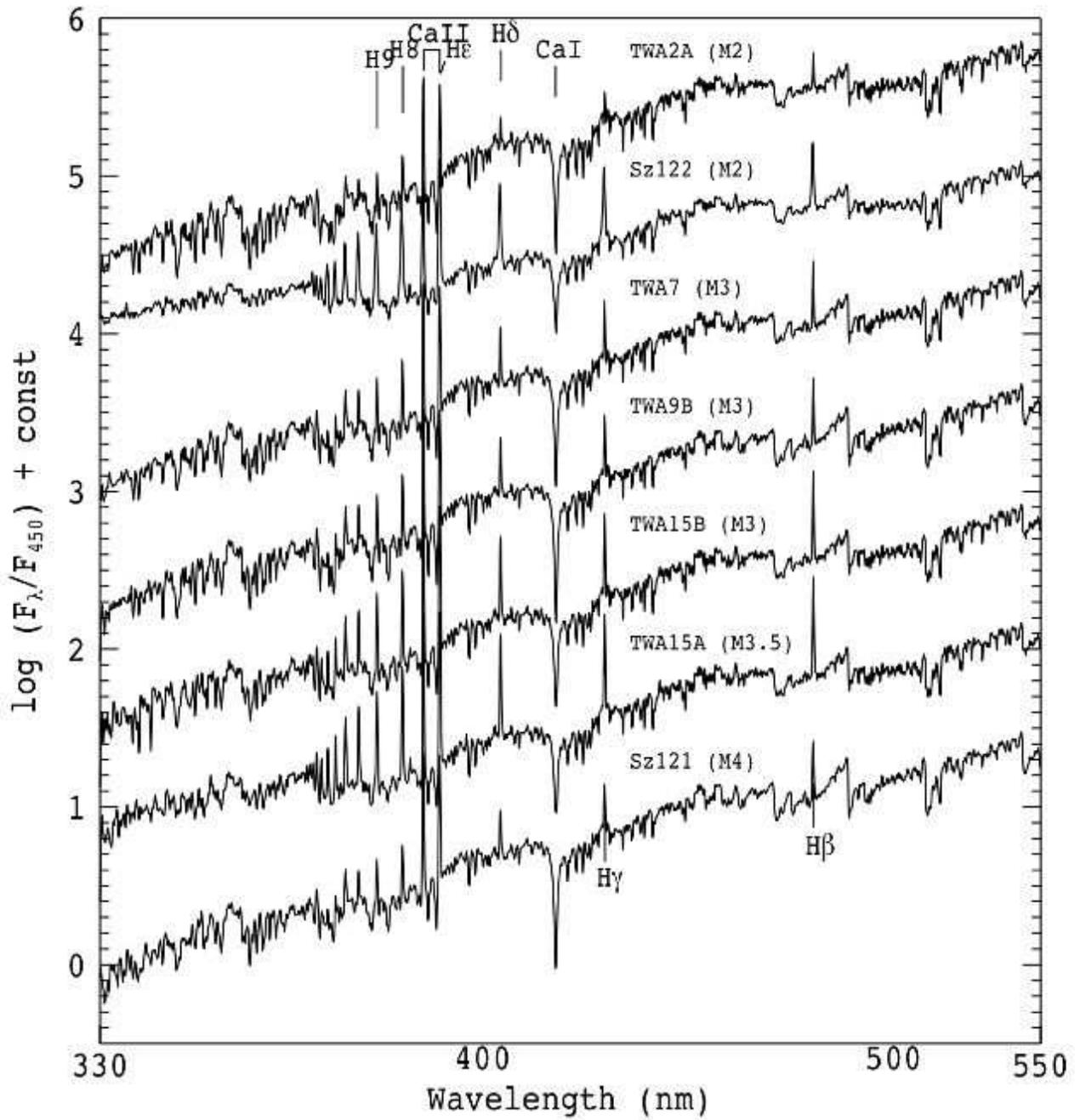}
      \caption{Same as Fig.~\ref{spec2_UVB}, but for spectral types between M2 and M4. }
	\label{spec2_UVB}
   \end{figure*}

   \begin{figure*}[!]
   \centering
   \includegraphics[width=\textwidth]{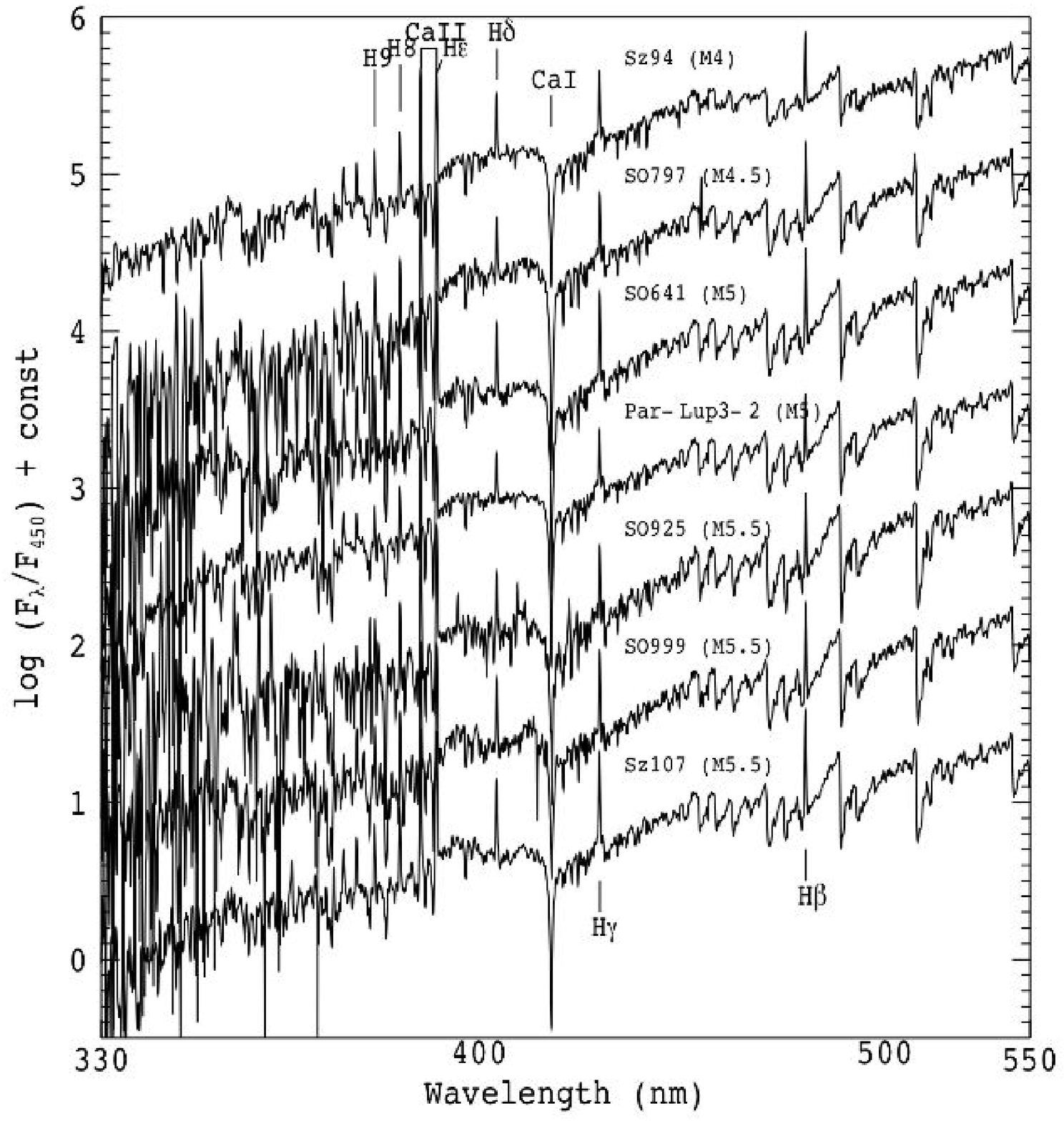}
      \caption{Same as Fig.~\ref{spec3_UVB}, but for spectral types later than M4. }
	\label{spec3_UVB}
   \end{figure*}

\end{document}